\documentstyle[aps,pra,epsfig]{revtex}

\newfont{\fds}{cmtt10 scaled 1000}
\newcommand{\ds}{
\mbox{\hspace*{-0.1ex}.\hspace*{-0.3ex}{\fds *}\hspace*{0.1ex}}}
\newcommand{\be}{\begin{equation}}
\newcommand{\ee}{\end{equation}}
\newcommand{\bea}{\begin{eqnarray}}
\newcommand{\eea}{\end{eqnarray}}
\def\P{{\cal P}}
\def\C{{\cal C}}

\def\E{{\cal E}}
\def\F{{\cal F}}
\def\O{{\cal O}}

\newcommand{\bra}[1]{\mbox{$\langle #1 |$}}
\newcommand{\ket}[1]{\mbox{$| #1 \rangle$}}

\def\>{\rangle}
\def\<{\langle}
\def\dag{\dagger}
\def\ot{\otimes}
\def\tr{{\rm tr}}
\def\non{\nonumber}
\def\noin{\noindent}
\newcommand{\eq}[1]{Eq.~(\ref{eq:#1})}

\def\lpm{ \left(\rule{0pt}{2.1ex}\right. }
\def\rpm{ \left.\rule{0pt}{2.1ex}\right) }
\def\lbL{ \left[\rule{0pt}{2.4ex}\right. }
\def\rbL{ \left.\rule{0pt}{2.4ex}\right] }
\def\lpL{ \left(\rule{0pt}{2.4ex}\right. }
\def\rpL{ \left.\rule{0pt}{2.4ex}\right) }

\def\laL{ \left|\rule{0pt}{2.4ex}\right. }
\def\raL{ \left.\rule{0pt}{2.4ex}\right| }
\newcommand{\mattwo}[4]{\left(
        \begin{array}{rr}{#1}&{#2}\\{#3}&{#4}\end{array}\right)}
\newcommand{\matthree}[9]{\left[
        \begin{array}{rrr}{#1}&{#2}&{#3}\\
                          {#4}&{#5}&{#6}\\
                          {#7}&{#8}&{#9}
                          \end{array}\right]}

\begin{document}
\draft

\title{Optimal simulation of two-qubit Hamiltonians\\
using general local operations}

\author{
C.~H.~Bennett,$^1$
J.~I.~Cirac,$^2$
M.~S.~Leifer,$^3$
D.~W.~Leung,$^{1*}$
N.~Linden,$^3$
S.~Popescu,$^4$
G.~Vidal$^2$\footnote{Correspondence:
wcleung@watson.ibm.com and Guifre.Vidal@uibk.ac.at.}}


\address{\vspace*{1ex}
$^1$IBM TJ Watson Research Center, P.O. Box 218, Yorktown Heights, NY
10598, USA \\
$^2$Institut f\"ur Theoretische Physik, Universit\"at Innsbruck,
A-6020 Innsbruck, Austria \\
$^3$Department of Mathematics, University of Bristol, University Walk,
Bristol, BS8 1TW, UK \\
$^4$H.H.~Wills Physics Laboratory, University of Bristol, Tyndall
Avenue, Bristol, BS8 1TL, UK, and \\ BRIMS, Hewlett-Packard
Laboratories, Stoke Gifford, Bristol BS12 6QZ, UK
\vspace*{1ex}}


\date{\today}


\maketitle


\begin{abstract}

We consider the simulation of the dynamics of one nonlocal
Hamiltonian by another, allowing arbitrary local resources but no
entanglement nor classical communication.
We characterize notions of simulation, and proceed to focus on
deterministic simulation involving one copy of the system.
More specifically, two otherwise isolated systems $A$ and $B$
interact by a nonlocal Hamiltonian $H \neq H_A+H_B$.  We consider
the achievable space of Hamiltonians $H'$ such that the evolution
$e^{-iH't}$ can be simulated by the interaction $H$ interspersed
with local operations.
%
%
For any dimensions of $A$ and $B$, and any nonlocal Hamiltonians
$H$ and $H'$, there exists a scale factor $s$ such that for all
times $t$ the evolution $e^{-iH'st}$ can be simulated by $H$
acting for time $t$ interspersed with local operations.
For 2-qubit Hamiltonians $H$ and $H'$, we calculate the optimal
$s$ and give protocols achieving it.
The optimal protocols do not require local ancillas, and can be
understood geometrically in terms of a polyhedron defined by a
partial order on the set of 2-qubit Hamiltonians.
\end{abstract}


\section{Introduction}

\subsection{Motivation}

Like the mythical lovers Thisbe and Pyramus, Alice and Bob wish to be
forever in each other's company, a situation described physically by
some many-atom interaction Hamiltonian $H'$.  Unfortunately their
parents disapprove, and have built a massive wall to keep the
youngsters apart. Fortunately there is a small hole in the wall, just
big enough for one atom of Alice to interact with one atom of Bob via
the two-atom interaction Hamiltonian $H$ (Fig.~1).
Can they use this limited interaction, together with local operations
on each side of the wall, to simulate the desired interaction $H'$?
Yes, if they are patient, because any nontrivial bipartite interaction
can be used both to generate entanglement and to perform classical
communication.  Therefore they can use $H$, along with local ancillary
degrees of freedom on each side of the wall, to generate enough
entanglement, and perform enough classical communication to teleport
Alice's entire original state to Bob's side. Now that they are
(virtually) together, Alice and Bob can interact to their heart's
content.  When it is time for Alice to go home, they teleport her back
to her side, in whatever entangled state they have gotten themselves
into, again using $H$ to generate the needed entanglement and perform
the needed classical communication.  So, by the time they get to be
old lovers, Alice and Bob can experience exactly what it would have
been like to be young lovers, if they are still foolish enough to want
that.

\begin{figure}
\centerline{\mbox{\psfig{file=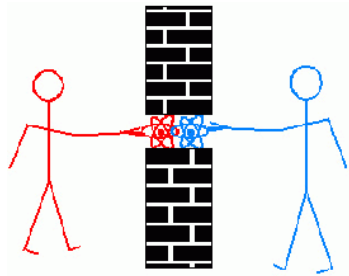,width=1.8in}}}
\caption{Thisbe (L) and Pyramus, separated by a wall, through
which they can only interact by a $2$-atom Hamiltonian $H$.}
\label{fig:pt}
\end{figure}

A more practical motivation for studying the ability of nonlocal
Hamiltonians to simulate one another comes from quantum control
theory~\cite{QuantumControl}, in particular the problem of using an
experimentally available interaction, together with local operations,
to simulate the evolution that would have occurred under some other
Hamiltonian not directly accessible to experiment.  A more
mathematical motivation comes from the desire to parameterize the
nonlocal properties of interaction Hamiltonians, so as to characterize
the efficiency with which they can be used to simulate one another,
and perform other tasks like generating entanglement
\cite{Dur00,ZZF00} or performing quantum
computation~\cite{Linden98,Leung99,Jones99,Dodd01}.  This parallels
the many recent efforts to parameterize the nonlocal properties of
quantum states, so as to understand when, and with what efficiency,
one quantum state can be converted to another by local operations, or
local operations and classical communication. It is not difficult to
see, by the Pyramus and Thisbe argument, that all nonlocal
Hamiltonians are {\em qualitatively\/} equivalent, in the sense that
for any positive $t'$ and $\epsilon$, there is a time $t$ such that
$t'$ seconds of evolution under $H'$ can be simulated, with fidelity
at least $1\!\!-\!\epsilon$, by $t$ seconds of evolution under $H$,
interspersed with local operations; but much work remains to be done
on the {\em quantitative\/} efficiency of such simulations.

In this paper we derive bounds on the time efficiency with which
one Hamiltonian can simulate another using local resources.
In the case of two interacting qubits, we show that these bounds are
optimal.
The structure of the paper is as follows.  In Sec.~\ref{sec:sframe},
we define the allowed resources and the type of simulation we consider. 
In Sec.~\ref{sec:example}, we prove some general results on the type
of simulation we consider along with some examples.  In
Sec.~\ref{sec:qna}, we define our goal and summarize our main results
for two-qubit Hamiltonians, that are proved in Secs.~\ref{sec:lu2} and
\ref{sec:luanc}.  Some discussions and conclusions, and more auxiliary
results can be found in Sec.~\ref{sec:discussion}, \ref{sec:conclusion} 
and Apps.~\ref{sec:simnots}-\ref{sec:derivelinform}.
We first describe in more detail some related results.

\subsection{Related work}
\label{sec:relwork}

The qualitative equivalence of nonlocal Hamiltonians noted above,
and the use of interaction as an infinitesimal generator of
entanglement, was already noted several years
ago~\cite{prehistory}.
These discussions also considered the question of interconverting
discrete nonlocal primitives, such as nonlocal gates, shared EPR
pairs, and uses of a classical bit channel.
More generally and quantitatively one may ask, given a nonlocal
Hamiltonian $H_{AB}\neq H_A+H_B$, what is the optimal efficiency
with which it can be used, in conjunction with local operations,

\begin{itemize}
\item to generate entanglement between $A$ and $B$
\item to transmit classical or quantum information from $A$ to $B$, 
or vice versa
\item to simulate the operation of another nonlocal Hamiltonian $H'$.
\end{itemize}
A partial answer to the first question, for two-qubit
Hamiltonians, was given by Ref.~\cite{Dur00}.
The current work is a continuation of previous efforts to study
the efficiency simulating one Hamiltonian by another.

Hamiltonian simulation has been considered in the context of
quantum computation~\cite{Linden98,Leung99,Jones99,Dodd01}
\cite{Wocjan01,Leung01,Stollsteimer01}.
%
%
In these works the system consists of $n$ qubits, with some given
{\em pairwise} interaction Hamiltonian.
In Refs.~\cite{Linden98,Leung99,Jones99}, the given Hamiltonian
was a sum of $\sigma_z \otimes \sigma_z$ interaction terms between
distinct qubits (see Sec.~\ref{sec:ineffsim} for definitions) and
the goal was to simulate a particular one of these terms.
This was extended in Refs.~\cite{Dodd01,Leung01,Stollsteimer01} to
arbitrary pairwise interactions, in both the simulating and the
simulated Hamiltonians.
In these papers the main concern was to obtain methods for
simulation, and therefore upper bounds on the resources as a
function of $n$.

Independent results on optimizing the time used of a given
Hamiltonian for performing certain tasks are reported in
Refs.~\cite{Wocjan01,Janzing01,Khaneja01}.
Reference~\cite{Wocjan01} gives a necessary condition for simulating
one $n$-qubit pairwise interaction Hamiltonian by another, and gives a
necessary and sufficient condition for simulation with a particular
given Hamiltonian. 
Time resources for simulating the inverse of a Hamiltonian are
discussed in Refs.~\cite{Wocjan01,Leung01,Janzing01}.
Reference~\cite{Khaneja01} considers simulating a unitary gate
using a given Hamiltonian and a set of controllable gates in the
shortest time.  A general framework is set up in terms of
Riemannian geometry. A time optimal protocol is obtained for the
specific Hamiltonian $\sigma_z \otimes \sigma_z$ in the $2$-qubit
case.

Finally, some more recent results have appeared since the original
posting of this preprint, extending it and related work in various
ways \cite{Janzing01sa,Janzing01sb,Vidal01sa,Vidal01sb} 
\cite{Wocjan01sa,Nielsen01sa,Wocjan01sb,Chen01s}.

\section{Simulation framework}
\label{sec:sframe}

In this section we describe our framework of Hamiltonian
simulation, i.e.~the rules under which the simulation is to be
performed.  We also describe other possible frameworks and their
relations to the one we adopt.

\subsection{Available resources}
\label{sec:resources}

Let $H$ and $H'$ each be a nonlocal Hamiltonian acting on two
isolated systems $A$ and $B$.
We consider the problem of simulating $H'$ by $H$ using unlimited
local resources.
These include instantaneous local operations and uncorrelated local
ancillas of any finite dimensions.
It is also necessary to allow some initial classical correlation
-- Alice and Bob are assumed to have agreed beforehand on their
time and spatial coordinates and the simulation protocol to be
followed.
Besides this, no other nonlocal resources are allowed, neither
prior entanglement nor any form of communication beyond what can
be achieved through the interaction $H$ itself.
Our goal is to minimize the time required of the given Hamiltonian
$H$ to simulate another Hamiltonian $H'$.  This will be defined
more formally in Sec.~\ref{sec:qna}.

Note that either the simulating or the simulated system or both can be
given the freedom of bringing in {\em local} degrees of freedom
(ancillas) and allowing interaction between each ancilla with the
corresponding local system.  Ancillas on the simulated system can make
the simulated more powerful and therefore harder to simulate.
Ancillas on the simulating system potentially make the simulation
easier.  We will allow ancillas on the simulating system, though 
they may not always help (Section \ref{sec:luanc}).

\subsection{One-shot and deterministic simulations} 
\label{sec:determ}

In this paper we only concern ourselves with protocols that are
one-shot---i.e.~operate on a single copy each of the simulated and
simulating systems---and that are required to succeed with probability
$1$. \vspace*{1ex} 

More generally, a simulation can be ``blockwise'', in which
$H^{\otimes n}$ is used for the simulation of $H'^{\otimes n}$, or in
which $H$ is time-shared among many copies of the system and the
amortized cost is considered.  A simulation can also be stochastic and
fail with finite probability, in which case the expected cost is
considered.

\subsection{Gate versus dynamics simulations}
\label{sec:dynamics}

One possible notion of simulation is that, given $H'$ and $t'$, we
simulate the final unitary evolution $e^{-iH't'}$ by composing
local operations with elements in the one-parameter family
$\{e^{-iHt}\}_t$.\footnote{ The evolution due to a Hamiltonian $H$
is given by $e^{-iHt}$. Note the $-$ minus in the exponent. }
The final evolution needs to be correct, but the intermediate
evolution need not correspond to $e^{-iH't''}$ for $0 \leq t'' <
t'$. The efficiency, given by the ratio $t/t'$ can depend on $t'$.
For example, $H$ can be used to generate entanglement and
classical communication to bring $A$ and $B$ together by
teleportation, apply $e^{-iH't'}$, and teleport $A$ back.  Viewing
the cost $t$ as a function of $t'$, $t$ does not increases
indefinitely with $t'$, rather, it can be made constant after it
reaches a sufficiently large value.
As another example, if the nonlocal Hamiltonian $H' = \sigma_z \ot
\sigma_z$ acts for time $t'=\pi/2$, the result is the unitary gate $i
\sigma_z \ot \sigma_z$, which is local, and requires no nonlocal
interaction time at all to simulate.
This type of simulation, with very different primitives, is much
studied in the context of universality of quantum gates (composing
a small set of available gates to obtain any desired unitary
gate).
More recently, simulation of a unitary gate using a fixed given
Hamiltonian for a minimal amount of time and local manipulations
was studied in Ref.~\cite{Khaneja01} and some partial results were
obtained.
{From} now on, we call this type of simulation ``gate simulation''
or ``finite time simulation''.
\vspace*{1ex}

A natural direction to strengthen the above notion of Hamiltonian
simulation is to require not only the end result, but also the
intervening {\em dynamics} of $H'$ to be simulated.  
Intuitively, one might expect this to mean that the application of
$H$, interspersed with instantaneous local operations, produces a
trajectory that remains continuously close to the trajectory $e^{-iH't}$
which one wishes to simulate.
However, this is impossible in general, because the needed local
operations cause the simulating trajectory to be discontinuous,
agreeing only intermittently with the trajectory one wishes to
simulate.
Accordingly we adopt the following definition of
dynamics simulation: 
\ The Hamiltonian $H$ simulates the dynamics of $H'$ with efficiency
$\mu$ if $\forall_{t'>0}$, $\forall_{\epsilon>0}$ the unitary operation
$e^{-i H' t'}$ can be simulated with fidelity $\ge 1\!-\!\epsilon$ 
by some protocol using $H$ for a total time $t'/\mu$ and local
operations.  
While this characterization may appear to have given up the idea of
approximating the simulated system at intermediate times, in fact it
has not, because it can be shown to imply the existence of a
$\mu$-efficient ``stroboscopic'' simulation, which approximates the
simulated trajectory arbitrarily closely not only at the begining and
end, but also at an arbitrary large set of intermediate times. 
We discuss this and other simulation notions in
Appendix~\ref{sec:simnots}.  We also show that the existence of a
protocol for dynamics simulation is equivalent to the existence of one
for simulating an infinitesimal time (see Sec.~\ref{sec:inftind})
which in turns implies the ability to create protocols for arbitrary
finite times by appropriately rescaling and repeating the
infinitesimal-time protocol. (see Appendix.~\ref{sec:derivelinform}).

%
%
%
%
%
%

\section{General results and examples}
\label{sec:example}

Having defined the simulation framework, we derive some important
general results and provide some examples of dynamics simulation,
which motivate our main results and simplify some of the later
discussions.

\subsection{Infinitesimal and time independent simulation}
\label{sec:inftind}

First of all we show that dynamics simulation is equivalent to
``infinitesimal simulation'', the problem of simulating the evolution
of $H'$ for an infinitesimal amount of time $t'$.
On one hand, any protocol for dynamics simulation simulates the
initial evolution, therefore is a protocol for infinitesimal
simulation.  On the other hand, iterating an infinitesimal simulation
results in dynamics simulation.
We restrict our attention to infinitesimal simulation from now on, and
focus on the lowest order effects in $t'$.
Note that this property may not hold for other types of simulation
described in Appendix~\ref{sec:simnots}. 
\vspace*{1ex}

Infinitesimal simulation has a very special structure -- the
optimal simulation protocol is independent of the infinitesimal
value of $t'$. The proof is included in Appendix
\ref{sec:derivelinform}.

\subsection{Local Hamiltonians are irrelevant}
\label{sec:local}

A general bipartite Hamiltonian $K$ can be written as,
\be
    K = K_A \ot I + I \ot K_B
      + \sum_{ij} M_{ij} \, \eta_i \ot \eta_j \,,
\ee
where $I$ denotes the identity throughout the paper, $K_A$, $K_B$
are local Hamiltonians acting on $A$, $B$ respectively, and
$\{\eta_i\}$ is a basis for traceless hermitian operators acting
on each of $A$ and $B$.
We can ``dispose'' of the local Hamiltonians $K_A$ and $K_B$ by
undoing them with local unitaries on $A$ and $B$:
\be
    (e^{i K_A t}  \ot  e^{i K_B t}) \; e^{-i t K} =
    e^{-i \, (K - K_A \ot I  - I \ot K_B) \, t} + {\cal O}(t^2)
\,.
\label{eq:localsim}
\ee
In other words, $K$ can be made to simulate its own nonlocal
component.
\vspace*{1ex}

Likewise, any Hamiltonian can simulate itself with additional
local terms.  Therefore, given unlimited local resources, the
problem of simulating an arbitrary Hamiltonian $H'$ by another
arbitrary one $H$ reduces to the case when both are purely
nonlocal.

\subsection{Possible inefficiencies in simulation}
\label{sec:ineffsim}

Consider the simplest case of two-qubit systems.  We introduce the
Pauli matrices
\be
    \sigma_x = \mattwo{0}{1}{1}{0} \,,~~
    \sigma_y = \mattwo{0}{-i}{i}{0} \,,~~
    \sigma_z = \mattwo{1}{0}{0}{-1} \,,~~
\label{pauli}
\ee
and the useful identity
\be
    U e^{M} U^\dag = e^{UMU^\dag}
\label{eq:cbasis}
\ee
where $M$ is any bounded square matrix and $U$ is any unitary
matrix of the same dimension.
\vspace*{1ex}

As an example, let $H = \sigma_x \ot \sigma_x$ and $H' = {1 \over
3} (\sigma_x \ot \sigma_x + \sigma_y \ot \sigma_y + \sigma_z \ot
\sigma_z)$.
To simulate $H'$ by $H$, let $U_1 = {1 \over \sqrt{2}}(\sigma_x +
\sigma_y)$ and $U_2 = {1 \over \sqrt{2}}(\sigma_x + \sigma_z)$, so
that $\sigma_y = U_1 \sigma_x U_1^\dag$ and $\sigma_z = U_2
\sigma_x U_2^\dag$.  Using \eq{cbasis}, it is easily verified that
\be
 e^{-i H' t'} = \lpm e^{-i H t'/3} \rpm
 \; \times \; \lpm U_1 \ot U_1 \; e^{-i H t'/3} \; U_1^\dag \ot U_1^\dag \rpm
 \; \times \; \lpm U_2 \ot U_2 \; e^{-i H t'/3} \; U_2^\dag \ot U_2^\dag \rpm
\,.
\ee
Conversely, we can simulate $H$ with $H'$:
\be
    e^{-i H t} = e^{-i H' 3t/2} \; \times \; \lpm
        \sigma_z \ot I \,\; e^{-i H' 3t/2} \,\; \sigma_z \ot I \rpm
\ee
Note that the simulating of $H'$ for a duration of $t'$ requires
applying $H$ for a duration of $t'$ whereas simulating $H$ for a
duration $t$ requires applying $H'$ for a duration of $3t$.
As the time required of the given Hamiltonian is a resource to be
minimized, we see that some simulations are less efficient than
the others.
In this paper, we are concerned with the inefficiencies of
simulation intrinsic to the Hamiltonians $H$ and $H'$ that are not
caused by a bad protocol.
For example, we will show later that the inefficiency in the above
example is intrinsic.


\subsection{Simulating the zero Hamiltonian -- stopping the evolution}
\label{sec:decouple}

In some applications, the given Hamiltonian $H$ cannot be switched
on and off.  Simulating the zero Hamiltonian ${\bf 0}$ can be
viewed as a means for switching off the Hamiltonian
$H$~\cite{Linden98,Leung99,Jones99}. This can always be done for
any dimensions of $A$ and $B$.

First, let $A$ and $B$ be $2^n$-dimensional, and
\be
    H = \sum_{\bf ij} c_{\bf ij} P_{\bf i} \ot P_{\bf j}
\,,
\label{eq:nqubitham}
\ee
where ${\bf i}$ is a binary vector $(i_1,i_2,\cdots,i_{2n})$ that
labels the $n$-qubit Pauli matrix $P_{\bf i} = \sigma_x^{i_1}
\sigma_z^{i_2} \ot \cdots \ot \sigma_x^{i_{2 \hspace*{-0.15ex}
n\!-\!1}} \sigma_z^{i_{2n}}$.
It is easily verified that
\be
    {1 \over 2^{2n}}
    \sum_{\bf i} P_{\bf i} M P_{\bf i} = \tr M \; {I \over 2^n}
\,.
\label{eq:decouple}
\ee
A protocol for simulating ${\bf 0}$ by $H$ is given by,
\be
    \Pi_{\bf ij} \, (P_{\bf i} \ot P_{\bf j}) \, e^{-i H t/2^{4n}} \,
                (P_{\bf i}^\dag \ot P_{\bf j}^\dag)
    ~=~
    e^{-{it \over 2^{4n}} \sum_{\bf ij} \, (P_{\bf i} \ot P_{\bf j}) \,
            H \, (P_{\bf i}^\dag \ot P_{\bf j}^\dag)  + {\cal O}(t^2)}
    ~\approx~ e^{-i \, t \, \tr H /2^{2n}}
\,,
\ee
in which the net evolution is just an overall phase to the lowest
order in $t$.

When $A$ and $B$ are $d$-dimensional, one can embed each of
$A$ and $B$ in a larger, $2^n$-dimensional system for $n = \lceil
\log_2 d \rceil$ to perform the simulation.
Physically, this can be done on each of $A$ and $B$, by attaching
a qubit ancilla, extending the Hilbert space to $2d$-dimensions,
and applying the simulation to a $2^n$-dimensional subspace, such
as one spanned by $|i\>\ot|0\>$ for $i=1,\cdots,d$ and
$|i\>\ot|1\>$ for $i=1 \cdots 2^n\!-d$.
Such simulation can also be done without ancillary degrees of
freedom, and an alternative method based on Ref.~\cite{Viola99} is
given in Appendix \ref{sec:viola}.

\subsection{Arbitrary but inefficient simulations}
\label{sec:anysim}

We now show that any nonlocal bipartite Hamiltonian can be used to
simulate any other, albeit with inefficiencies.
In other words, for any $H$ and $H'$, operating $H$ for time $t$
can simulate the evolution of $H'$ for time $t'$ with $t'/t > 0$.
This holds for any dimensions.
We keep all definitions from the previous example in the following
protocol.
\vspace*{1ex}

First, let $A$ and $B$ be $2^n$-dimensional, $H = \sum_{\bf ij}
c_{\bf ij} P_{\bf i} \ot P_{\bf j}$ and $H' = \sum_{\bf ij} c_{\bf
ij}' P_{\bf i} \ot P_{\bf j}$.
Without loss of generality the coefficient for $P_{\bf k} \ot
P_{\bf k}$ is positive, i.e. $c_{\bf kk} > 0$, where ${\bf k} =
(0,1,0,\cdots,0)$ and $P_{\bf k} = \sigma_z \ot I \ot \cdots \ot
I$.
It is known that for any $P_{\bf i}$ and $P_{\bf j}$, there exist
unitary operations $U_{{\bf ij}\pm}$ in the {\em Clifford
group}~\cite{Gottesman97}, such that
\be
    U_{{\bf ij}\pm} P_{\bf i} U_{{\bf ij}\pm}^\dag = \pm P_{\bf j}
\label{eq:clifford}
\ee
In other words, one can always transform any $P_{\bf i}$ to any
other or to its negation.
In our protocol, $H$ simulates $H'$ in two steps.  First, $H$
simulates $P_{\bf k} \ot P_{\bf k}$ by:
\be
    \Pi_{{\bf i,i'}|i_1,i'_1=0} \; P_{\bf i} \ot P_{\bf i'}
    \; e^{-i H t/2^{4n\!-\!2}} P_{\bf i} \ot P_{\bf i'}
    ~\approx~ e^{-{i t \over 2^{4n\!-\!2}} \sum_{{\bf i,i'}|i_1,i'_1=0} \!
    P_{\bf i} \ot P_{\bf i'} H P_{\bf i} \ot P_{\bf i'}}
    ~=~ e^{-i \, t \, c_{\bf kk} P_{\bf k} \ot P_{\bf k}
    \, + \, {\rm local~terms}}
\ee
Alice and Bob independently apply an averaging over all Pauli
operators commuting with $P_{\bf k}$, removing all operators
except for $I = P_{\bf 0}$ and $P_{\bf k}$ in each of their
systems. The local terms can be ignored, following
Sec.~\ref{sec:local}.
Second, $P_{\bf k} \ot P_{\bf k}$ simulates $H'$ by:
\be
    \Pi_{\bf ij}
    \lpm
    U_{{\bf ki} \; {\rm sg}(\!c_{i\!j}'\!)} \ot U_{{\bf kj}+}
    \rpm \;
    e^{-i P_{\bf k} \ot P_{\bf k} |c_{ij}'| \: t'} \;
    \lpm
    U_{{\bf ki} \; {\rm sg}(\!c_{i\!j}'\!)} \ot U_{{\bf kj}+}
    \rpm^\dag
    ~\approx~ e^{-i t' \sum_{\bf ij} P_{\bf i} \ot P_{\bf j} c_{ij}'}
    ~=~ e^{-i H' t'}
\ee
where ${\rm sg}(x) = x/|x|$ if $x \neq 0$ and we omit terms with
$c_{ij}'=0$ .
\vspace*{1ex}

When $A$ and $B$ are $d$-dimensional, the simulation of $sH'$ by
$H$ can again be performed in a larger $2^n \times 2^n$ system.
This method implies a lower bound on the maximum possible value of
$s$, $s \geq {1 \over 2^{2 \lceil \log_2 d \rceil}} {\max_{\bf ij}
|c_{\bf ij}| \over \sum_{\bf ij} |c_{\bf ij}'|}$.
It is also possible to perform the simulation without ancillas.
The proof is given in Appendix \ref{sec:ddimsim}.  Other methods
for such simulation were independently reported in
\cite{Wocjan01sa,Nielsen01sa,Wocjan01sb}.

\subsection{Equivalent classes of local manipulations}
\label{sec:eqc}

Under our simulation framework, Alice and Bob are given unlimited
local resources.  In this subsection, we show that they only need
a relatively small class of manipulations.  To facilitate the
discussion, we introduce classes of operations $C$, that can be
LU, LO, LU$+$anc, and LO$+$anc, to be defined as follows.
LU is the class of all local unitaries that act on $A \otimes B$.
LU$+$anc is similar, but acts on $(A \ot A') \ot (B \ot B')$ where
$A'$ and $B'$ are uncorrelated ancillary systems of any finite
dimension.  LO and LO$+$anc are similarly defined, with the
unitaries replaced by general trace-preserving quantum operations.
Note that the largest class LO$+$anc corresponds to what is most
generally allowed under our simulation framework.
\vspace*{1ex}

We now show that LU$+$anc, LO, and LO$+$anc are equivalent under
our framework.
First, we show that LU$+$anc is at least as powerful as LO$+$anc.
Any trace preserving quantum operation can be implemented by
performing a unitary operation on a larger Hilbert space, followed
by discarding the extra degrees of freedom~(see, for example,
Ref.~\cite{Nielsen00}).
The exact difference between LO$+$anc and LU$+$anc is that
measurements and tracing are disallowed in the latter.
However, these are not needed when simulating Hamiltonian in
LU$+$anc, due to the following facts.
(1) Measurements can be delayed until the end of the protocol, as
operations conditioned on intermediate measurement results can be
implemented unitarily.
(2) In Hamiltonian simulation, the ancillary systems $A'B'$ have
to be disentangled from $AB$ at the end of the simulation. Thus no
actual measurement or discard is needed.
These facts allow any LO$+$anc protocol to be reexpressed as an
LU$+$anc protocol with pure product state ancillas, meaning that
LO and LO$+$anc are no more powerful than LU$+$anc.
Conversely, due to fact (2) above, any LU$+$anc protocol can be
viewed as an LO protocol.
Thus, we establish the equivalence between LO, LU$+$anc, and
LO$+$anc. {From} now on, we focus on LU$+$anc protocols for full
generality, and on LU protocols as a possible restriction.

\section{Formal statement of the problem and summary of results}
\label{sec:qna}

Let $H$, $H'$, $A$, $B$, $A'$, $B'$ be defined as before.

\begin{quote}
{\bf Definition:} $H'$ can be {\em efficiently simulated} by $H$,
 \be
 H' \leq_C H,
 \ee
if the evolution according to $e^{-iH't'}$ for any time $t'$ can
be simulated by using the Hamiltonian $H$ for the same time $t'$
and using manipulations in the class $C$.
\end{quote}

\begin{quote}
{\bf Definition:} $H'$ and $H$ are {\em equivalent} under the
class $C$,
 \be
 H' \equiv_C H,
 \ee
if $H' \leq_C H$ and $H' \leq_C H$.
\end{quote}

Throughout the paper, we only consider LU$+$anc protocols
following Sec.~\ref{sec:eqc}.
We also restrict attention to $H$ and $H'$ that are purely
nonlocal, following Sec.~\ref{sec:local}.
%
\vspace*{1ex}

An LU$+$anc protocol simulates $H'$ with $H$ by interspersing the
evolution of $H$ with local unitaries on $AA'$ and $BB'$.
More specifically, the most general protocol for simulating $H'$
using $H$ for a total time $t$ is to attach the ancillas $A'B'$ in
the state $\ket{0_{A'}} \ot \ket{0_{B'}}$, apply some $U_1 \ot
V_1$, evolve $AB$ according to $H$ for some time $t_1$, apply $U_2
\ot V_2$, further evolve $AB$ according to $H$ for time $t_2$, and
iterate ``apply $U_i \ot V_i$ and evolve with $H$ for time $t_i$''
some $n$ times.  At the end, it applies a final $U_f \ot V_f$.  The
$t_i > 0$ are constrained\footnote{
Without loss of generality, a protocol with $\sum_{i=1}^n t_i < t$
can be turned to one with $\sum_{i=1}^n t_i = t$ by simulating the
zero Hamiltonian as described in Section \ref{sec:example}.} by
$\sum_{i=1}^n t_i = t$.
Suppose the protocol indeed simulates an evolution for time $t'$
according to $H'$.  Then we can write
\bea
       \lpL U_f \ot V_f \times U_n \ot V_n \; e^{-i H t_n} \;
    U_n^\dag \ot V_n^\dag
        \times \cdots \times
        U_1 \ot V_1 \; e^{-i H t_1} \;
        U_1^\dag \ot V_1^\dag \rpL \; |\psi\> \ot |0_{A'}\> \ot |0_{B'} \>
    \hspace*{20ex}
\non
\\
    \hspace*{50ex} = \lpm e^{-iH't'} |\psi\> \rpm
    \ot \lpm W_{A'B'}(t_1,\cdots,t_n) |0_{A'}\> \ot |0_{B'}\> \rpm
    \,,
\label{eq:gensim}
\eea
where we have redefined $U_{i=1,2,\cdots,n}$ and
$V_{i=1,2,\cdots,n}$, and $|\psi\>$ denotes the initial state in
$AB$.  In \eq{gensim}, $e^{-i H t_i}$ acts on $AB$ and implicitly
means $e^{-i H t_i} \ot I_{A'B'}$.  The operator
$W_{A'B'}(t_1,\cdots,t_n)$ describes the residual transformation
of $A'B'$, and can be chosen to be unitary since the operation on
the left hand side of \eq{gensim} is unitary.
%
%
The problem we are concerned with can be stated in two equivalent
ways:

\begin{quote}
{\bf Optimal and efficient simulation}: Let $H$ be arbitrary.  The
{\em optimal simulation problem} is to, for each $H'$, find a
solution $\{U_i\}$, $\{V_i\}$, $\{t_i\}$ of \eq{gensim} such that
$t'/t$ is maximal.  The {\em efficient simulation problem} is to
characterize every $H'$ which admits a solution for \eq{gensim}
with $t'=t$, i.e. $H' \leq_{{\rm LU}+{\rm anc}} H$.
\end{quote}

\begin{quote}
{\bf Definition}: The optimal simulation factor $s_{H'|H}$ under
class $C$ of operations is the maximal $s>0$ such that $sH' \leq_C
H$.
\end{quote}

The optimal and efficient simulation problems are equivalent
because inefficient simulation is always possible (see Section
\ref{sec:example}).  The efficient simulation problem can be
solved by finding the optimal solution for each $H'$ and
characterizing those with $t'/t \geq 1$.  The optimal simulation
problem can be solved by finding the maximum $s$ for which $sH'$
is efficiently simulated. With this in mind, we may talk of
solving either problem throughout the paper.
\vspace*{1ex}


We now summarize our results.  We show in
Appendix~\ref{sec:derivelinform} that, in the infinitesimal
regime, the most general simulation protocol \eq{gensim} using
LU$+$anc is equivalent to
\be
    sH' =  \bra{0_{A'}}\otimes\bra{0_{B'}} \sum_i \;p_i
    \; U_i \ot V_i \; (H \ot I_{A'B'}) \; U_i^{\dagger} \ot V_i^{\dagger}
    \; \ket{0_{A'}}\otimes \ket{0_{B'}}
\,.
\label{eq:sumanc}
\ee
In the LU case (without ancillas), \eq{sumanc} reads
\be
    sH' = \sum_i \;p_i\;
    U_i \ot V_i \; H \; U_i^{\dagger} \otimes V_i^{\dagger}
\,,
\label{eq:posum}
\ee
where $t = t_1 + \cdots + t_n$, $p_k = t_k/t$, and $s = {t' \over
t}$.
Thus, the set $\{H' \leq_{\rm LU} H\}$ is precisely the convex
hull of the set $\{ U \ot V \; H \; U^\dag \ot V^\dag \}$ when $U$
and $V$ range over all unitary matrices on $A$ and $B$
respectively.
The linear dependence of ${t' \over t} H'$ on $H$ is manifest in
both \eq{sumanc} and \eq{posum}.
\vspace*{1ex}

Our main results apply to the simulation of two-qubit
Hamiltonians, and are summarized as follows:
\begin{quote}
{\bf Result 1}: {\em Any} simulation protocol using LU$+$anc can
be replaced by one using LU with the same simulation factor.  This
will be proved in Section~\ref{sec:luanc}.  Thus, the four partial
orders $\leq_{\rm LU}$, $\leq_{{\rm LU}+{\rm anc}}$, $\leq_{\rm
LO}$, $\leq_{{\rm LO}+{\rm anc}}$ are equivalent for two-qubit
Hamiltonians.

{\bf Result~2}: We present the necessary and sufficient conditions
for $H'\leq_{LU} H$, for arbitrary two-qubit Hamiltonians $H$ and
$H'$, and find the optimal simulation factor $s_{H'|H}$ and the
optimal simulation strategy in terms of $\{U_i\},\{V_i\},\{t_i\}$.
This will be discussed in Section~\ref{sec:lu2}.
\end{quote}

These results naturally endow the set of two-qubit Hamiltonians
with a partial order $\leq_C$.
This induces for each $H$, a set $\{H': H' \leq_C H\}$ which is
convex: if $H' \leq_C H$ and $H'' \leq_C H$, $p H'+ (1-p) H''
\leq_C H$ for any $0 \leq p \leq 1$.
Our method relies on the convexity of the set $\{H': H' \leq_C
H\}$, which has a simple geometric description, and in turns
allows the partial order $\leq_C$ to be succinctly characterized
by a majorization-like relation.  The geometric and majorization
interpretations offer two different methods to obtain, in
practice, the optimal protocol and the simulation factor.

\section{Optimal LU simulation of two-qubit Hamiltonians}
\label{sec:lu2}

We will prove that $\leq_{\rm LU}$ is equivalent to $\leq_{{\rm
LU}+{\rm anc}}$ in the next section.  In this section, we focus on
LU simulations.  We first adapt a result from Ref.~\cite{Dur00} to
reduce the problem to a smaller set of two-qubit Hamiltonians $H$
and $H'$.
Then, for any $H$, we identify the set $\{H': H' \leq_C H\}$ with
a simple polyhedron and obtain simple geometric and algebraic
characterizations of it.  The optimal solution for each pair of
$H$ and $H'$ is derived.  Finally, the problem is rephrased in the
language of majorization.

\subsection{Normal form for two-qubit Hamiltonians}
\label{sec:normform22}

The most general purely nonlocal two-qubit Hamiltonian $K$ can be
written as,
 \be
    K = \sum_{ij} M_{ij} \, \sigma_i \ot \sigma_j \,,
 \label{eq:K}
 \ee
where the summation is over Pauli matrices $i,j=x,y,z$ or $1,2,3$
throughout the discussion for two-qubit Hamiltonians.  Let
 \be
 H = \sum_i h_i \, \sigma_i \ot \sigma_i,
 \label{eq:H}
 \ee
where $h_1 \geq h_2 \geq |h_3|$ are the singular values of the $3
\times 3$ matrix $M$ with entries $M_{ij}$, and $h_3 = {\rm
sg}(\det M) |h_3|$.  We say $H$ is the {\em normal} form of $K$.

\begin{quote}
{\bf Theorem:} Let $H$ be the normal form of $K$.  Then $H
\equiv_{\rm LU} K$.

{\em Proof:} If the local unitaries $U^\dag \ot V^\dag$ and $U \ot
V$ are applied before and after $e^{-iKt}$, the resulting
evolution is given by
\be
    e^{-iK't} = U \ot V \; e^{-iKt} \; U^\dag \ot V^\dag
         = e^{-i \; (U \ot V) \; K (U^\dag \ot V^\dag ) \; t} \,,
\ee
with
\bea
    K' & = & (U \ot V) \; K \; (U^\dag \ot V^\dag )
\non
\\  & = & \sum_{ij} M_{ij} \;
        (U \sigma_i U^\dag) \ot (V \sigma_j V^\dag)
\non
\\  & = & \sum_{ij} M_{ij}
    \lpL \sum_l R_{il} \sigma_l \rpL \ot \lpL \sum_k S_{jk} \sigma_k \rpL
\label{eq:rotation}
\\  & = & \sum_{lk} \lpm R^T M S \rpm_{lk} \; \sigma_l \ot \sigma_k
    \; \equiv \; \sum_{lk} M'_{lk} \; \sigma_l \ot \sigma_k
\label{eq:Kprime}
\eea
In \eq{rotation}, $R,S\in$ SO(3) since conjugating $\vec{r} \cdot
\vec{\sigma}$ by SU(2) matrices corresponds to rotating $\vec{r}$
by a matrix in SO(3) (and vice versa).
Equation (\ref{eq:Kprime}) implies that $K' = U \ot V K U^\dag \ot
V^\dag$ for some unitary $U,V$ if and only if $M' = R^T M S$.
In particular, there is a choice of $R$ and $S$ that makes $K'=H$:
\be
    R^T = \left(\begin{array}{ccc}{1}&{0}&{0} \\ {0}&{1}&{0}
        \\ {0} & {0} & {\det O_1} \end{array} \right) \times O_1^T
    \,,~~~
    S = O_2^T \times
    \left(\begin{array}{ccc}{1}&{0}&{0} \\ {0}&{1}&{0}
        \\ {0} & {0} & \det O_2
    \end{array} \right)
\ee
where $M = O_1 D \, O_2$ is the singular value decomposition of
$M$, with $O_1, O_2 \in$ O(3) and $D = {\rm diag}(h_1, h_2,
|h_3|)$.
Thus $K$ and $H$ are related by a conjugation by local unitaries,
which implies $K \equiv_{\rm LU} H$.
\end{quote}

\noin As suggested by the above proof, we define a few useful
notations.
\begin{quote} {\bf Definitions}
We call the $3 \times 3$ real matrix $M_{ij}$ the ``Pauli
representation'' of $K$, when $M$ and $K$ are related by \eq{K}.
We use $D_K$ to denote a diagonal Pauli representation of $K$.
\end{quote}

\noin Since any 2-qubit Hamiltonian is equivalent to its normal
form, we assume $H'$, $H$ are in normal forms from now on.  We now
turn to LU simulation of $H'$ by $H$.

\subsection{General LU simulation of normal form two-qubit Hamiltonians}
\label{sec:glunorm}

Recall from \eq{posum} in Section \ref{sec:qna} that the most
general simulation using LU is given by
\be
     s H' =
     p_1 \; U_1 \ot V_1 \; H \; U_1^\dag \ot V_1^\dag + \cdots +
     p_n \; U_n \ot V_n \; H \; U_n^\dag \ot V_n^\dag
\label{eq:cosumH}
\ee
where $s = t'/t$.
Following the discussion in Section \ref{sec:normform22}, we only
need to consider $H = \sum_i h_i \; \sigma_i \ot \sigma_i$ and $H'
= \sum_i h_i' \; \sigma_i \ot \sigma_i$ that are in their normal
forms.
The Pauli representation of $(U \ot V) H (U^\dag \ot V^\dag)$ is
given by $R D_H S$ for some $R,S \in$ SO(3).
We can reexpress \eq{cosumH} as
\be
    s D_{H'} =  p_1 \; R_1 D_H S_1 + \cdots + p_n \; R_n D_H S_n
\label{eq:cosumM}
\ee
where $R_i,S_i \in SO(3)$.  Since $H$ and $H'$ are in their normal
form, $h_1 \geq h_2 \geq |h_3|$ and $h_1' \geq h_2' \geq |h_3'|$.
Without loss of generality, we can make two assumptions.
First, we can assume $h_3 \geq 0$: If $h_3 < 0$, we can
right-multiply \eq{cosumM} by $S = {\rm diag}(1,1,-1)$: \be
    s D_{H'} S =  p_1 \; R_1 (D_H S) (S S_1 S) + \cdots
    + p_n \; R_n (D_H S) (S S_n S)
\ee
in which $S S_i S \in$ SO(3), and $D_H S = {\rm
diag}(h_1,h_2,|h_3|)$ is of the desired form.  Thus, we can assume
$h_3 \geq 0$.
Second, note that $s_{H'|H} = a s_{H'|aH} = {1 \over a}
s_{aH'|H}$.
The protocol is unchanged when \eq{cosumM} is divided by $\tr D_H
= h_1 + h_2 + h_3$.  Therefore, without loss of generality, the
normalization $h_1 + h_2 + h_3 = 1$ can be assumed.
\vspace*{1ex}

Equations (\ref{eq:cosumH}) and (\ref{eq:cosumM}) have a simple
physical interpretation: the protocol partitions the allowed usage
of $H$ ($D_H$) into different $U_k \ot V_k \; H \; U_k^\dag \ot
V_k^\dag$ ($R_k D S_k$), resulting in an ``average Hamiltonian''
$H'$ ($D_{H'}$), which is a convex combination of the $U_k \ot V_k
\; H \; U_k^\dag \ot V_k^\dag$ ($R_k D S_k$).
\vspace*{1ex}

The Hamiltonians, represented by $D_{H'}$, that can be efficiently
simulated ($s=1$) correspond to the diagonal elements of the
convex hull of $\{R D_H S: R,S \in {\rm SO(3)}\}$.
We call this diagonal subset, which is also convex, $\C_H$.
Note that the zero Hamiltonian is in the {\em interior} of $\C_H$,
because $H$ can simulate any $sH'$ for small $s$ without ancillas
(see Section~\ref{sec:example}).
Thus $\forall~D_{H'} \neq 0$, the optimal solution is a {\em
boundary} point of $\C_H$.
The problem of efficient or optimal simulation can be rephrased:
\begin{quote}
Given $H$, let $\C_H$ be the diagonal subset of the convex hull of
$\{R D_H S : R,S \in {\rm SO(3)}\}$.  Then $H'$ can be efficiently
simulated by $H$ if and only if $D_{H'} \in \C_H$.
For any $H'$, $s_{H'|H} D_{H'}$, which represents the optimal
simulation, is the unique intersection of the semi-line $\lambda
D_{H'}$ ($\lambda \geq 0$) with the boundary of $\C_H$.
The optimal protocol can be obtained by decomposing $s_{H'|H} \,
D_{H'}$ in terms of the extreme points of $\C_H$.
\end{quote}

Since each point in $\C_H$ can be decomposed as a convex
combination of the extreme points of $\C_H$, each efficiently
simulated Hamiltonian can be identified with a simulation protocol
and vice versa.  We will refer to elements in $\C_H$ as
Hamiltonians or simulation protocols interconvertibly.
\vspace*{1ex}

Central to our problem is the structure of $\C_H$.  We will show
in Sec.~\ref{sec:ceqp} that it is a simple polyhedron that we call
$\P_H$.  Its set of vertices, $P_{24}$, is a subset of $\C_H$
containing $24$ elements.
%
%
They are obtained from $D_H$ by permuting the diagonal elements
and putting an even number of $-$ signs.
More explicitly, these elements are $\pi_{i} D_H \pi_i s_j$, where
\bea \non
    &\pi_0 = I \,,~&
    \pi_1 = \left[ \begin{array}{ccc}
        -1 & 0 & 0 \\
         0 & 0 & 1 \\
         0 & 1 & 0
        \end{array} \right] \,,~
    \pi_2 = \left[ \begin{array}{ccc}
         0 & 0 & 1 \\
         0 &-1 & 0 \\
         1 & 0 & 0
        \end{array} \right] \,,~
    \pi_3 = \left[ \begin{array}{ccc}
         0 & 1 & 0 \\
         1 & 0 & 0 \\
         0 & 0 & -1
        \end{array} \right] \,,~
    \pi_4 = \left[ \begin{array}{ccc}
         0 & 1 & 0 \\
         0 & 0 & 1 \\
         1 & 0 & 0
        \end{array} \right] \,,~
    \pi_5 = \left[ \begin{array}{ccc}
         0 & 0 & 1 \\
         1 & 0 & 0 \\
         0 & 1 & 0
        \end{array} \right] \,,~
\\
\non
    & s_0 = I \,,~ &
    s_1 = \left[ \begin{array}{ccc}
         1 & 0 & 0 \\
         0 & -1 & 0 \\
         0 & 0 & -1
        \end{array} \right] \,,~
    s_2 = \left[ \begin{array}{ccc}
        -1 & 0 & 0 \\
         0 & 1 & 0 \\
         0 & 0 & -1
        \end{array} \right] \,,~
    s_3 = \left[ \begin{array}{ccc}
        -1 & 0 & 0 \\
         0 & -1 & 0 \\
         0 & 0 & 1
        \end{array} \right] \,.
\eea
The transformation $D_H \rightarrow \pi_i D_H \pi_i^{\dagger} s_j$
is physically achieved by $H \rightarrow (U_{\pi_i}^\dag \!  \ot
U_{s_j} U_{\pi_i} )\, H \,( U_{\pi_i} \! \ot U_{s_j}^\dag
U_{\pi_i}^\dag)$, where $U_{\pi_i} = (\sigma_j +
\sigma_k)/\sqrt{2}$ for $i = 1,2,3$ and $i,j,k$ distinct,
$U_{\pi_i} = \cos(\pi/3) I \pm i \sin(2\pi/3) (\sigma_x + \sigma_y
+ \sigma_z)/\sqrt{3}$ for $i=4,5$, and $U_{s_i} = \sigma_i$ for
$i=1,2,3$.  These can be verified using \eq{rotation}.  The fact
that $P_{24}$ is the set of extreme points of $\C_H$ means that
any optimal simulation protocol only involves the transformations
$D_{H} \rightarrow \pi_i D_H \pi_i s_j$.
\vspace*{1ex}

In the next few subsections, we investigate the geometry of
$\P_H$, prove that $\C_H = \P_H$, and find the optimal solution
for any $H'$ using the fact $\C_H = \P_H$.  Then, we restate the
solution in terms of a majorization-like relation.

\subsection{The Polyhedron $\P_H$}
\label{sec:ph}

Since $P_{24}$ and $\P_H$ consist of diagonal matrices only, their
elements can be represented by real $3$-dimensional vectors.
The defining characterization of $\P_H$ is the polyhedron with
$24$ (not necessarily distinct) vertices that are elements of
$P_{24}$.
We now turn to a useful characterization of $\P_H$ as the region
enclosed by its faces,
\be
    (x,y,z) \in \P_H ~{\rm iff} \left\{
    \begin{array}{l}
    ~~|x| \leq h_1 ~, |y| \leq h_1 ~, |z| \leq h_1
\\  -(1-2 h_3) \leq +x+y+z \leq 1
\\  -(1-2 h_3) \leq -x-y+z \leq 1
\\  -(1-2 h_3) \leq +x-y-z \leq 1
\\  -(1-2 h_3) \leq -x+y-z \leq 1
    \end{array} \right.
\label{eq:ineq}
\ee
where the fact that $H$ is in normal form, $h_3 \geq 0$, and that
$h_1+h_2+h_3 = 1$ are used to replace the bounds $\sum_i h_i$ and
$-(\sum_i h_i - 2 \min_i h_i)$ by $1$ and $-(1-2 h_3)$ in \eq{ineq}.
Equation (\ref{eq:ineq}) can be used to determine whether a point,
as specified by its coordinates, is in $\P_H$ or not.  The
validity of \eq{ineq} can be proved by plotting $P_{24}$ (and
therefore $\P_H$) and verifying that the faces are as given in
\eq{ineq}.
We first plot $\P_H$ for the simple case $(h_1,h_2,h_3) =
(1,0,0)$, for which $P_{24}$ has $6$ distinct points: $(\pm
1,0,0)$, $(0,\pm 1,0)$, $(0,0,\pm 1)$ and \eq{ineq} holds
trivially:
\vspace*{-1ex} \be \mbox{\psfig{file=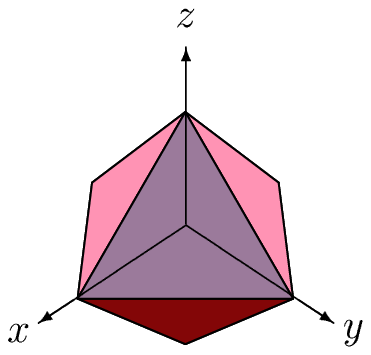,width=1.5in}}
\label{fig:octa}
\ee
Now, we plot $\P_H$ for the most complicated case, $h_1 > h_2 >
h_3 > 0$ in Fig.~(\ref{fig:polytope}):
\be \mbox{\psfig{file=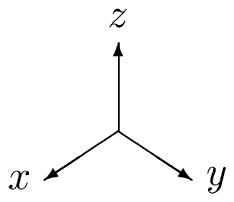,width=1.0in}} \hspace*{-7ex}
\mbox{\psfig{file=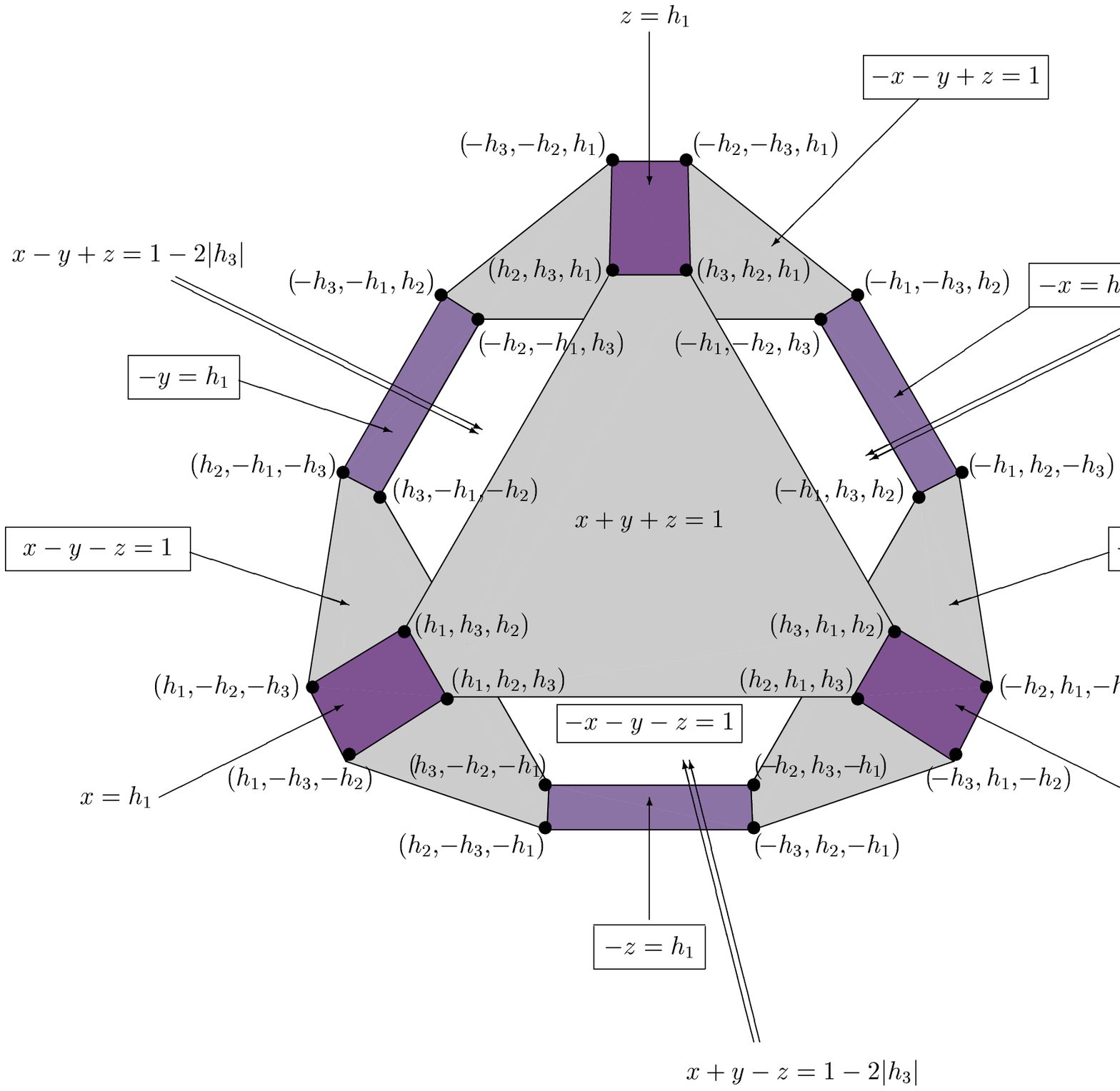,width=5.5in}}
\label{fig:polytope}
\ee

\noin As in Fig.~(\ref{fig:octa}), Fig.~(\ref{fig:polytope}) is
viewed from the direction $(1,1,1)$.
Three faces are removed to show the structure in the back.
There are $3$ types of faces.
There are $6$ identical rectangular purple faces on the planes
$x=\pm h_1, y = \pm h_1, z = \pm h_1$.
There are two groups each consists of $4$ identical hexagonal
faces.
The first group of $4$ consists of the $3$ light blue faces in the
back, and the light blue face in the front.  These are the
truncated faces of the original octahedron, lying on the planes
$x+y+z = 1, -x+y-z = 1, -x-y+z = 1, x-y-z = 1$.
The second group consists of the $3$ empty faces in the front, and
the white face in the back.  They are {\em inside} the original
octahedron and are parallel to the original faces.  They lie on
the planes $-x-y-z = 1-2 h_3, -x+y+z = 1-2 h_3, x-y+z = 1-2 h_3,
x+y-z = 1-2 h_3$.  Note that each hexagon in one group has a
parallel counterpart in the other group.
All together, there are $7$ pairs of parallel faces, each pair
bounds one expression in \eq{ineq}.
It is straightforward to verify the diagram and \eq{ineq}.
\vspace*{1ex}

The plots for other cases, such as when $h_3 = 0$ or $h_1 = h_2$,
can be likewise obtained and \eq{ineq} be verified.  These are
generally simpler then Fig.~(\ref{fig:polytope}), and may admit
simpler solutions in Section \ref{sec:opsol}.  However, we leave
the details to the interested readers and move on to prove that
$\C_H = \P_H$.

\subsection{Proof of $\C_H = \P_H$}
\label{sec:ceqp}

We now show that $\C_H = \P_H$.  Recall that $\C_H$ consists of
Hamiltonians that can be expressed as $D_{H'} = \sum_i p_i \; R_i
D_H S_i^T$ (by putting $s=1$ in \eq{cosumM} and using $S_i^T$ in
place of $S_i$).
The fact that $D_{H'}$ is diagonal implies that only the diagonal
elements in each $R_i D_H S_i^T$ contribute to $D_{H'}$; it is
possible for an individual $R_i D_H S_i^T$ to be off-diagonal, but
the off-diagonal elements have to cancel out in the sum.
To show that $\C_H = \P_H$, it suffices to show that the diagonal
part of each $R_i D_H S_i^T$ is in $\P_H$, because any $D_{H'} \in
\C_H$ will then be in $\P_H$.
\vspace*{1ex}

Let us consider the diagonal part of $R D_H S^T$, represented as a
$3$-dimensional vector $(g_1,g_2,g_3)$.  Since $D_H = {\rm
diag}(h_1,h_2,h_3)$,
\be
    g_i = (R D_H S^T)_{ii} = \sum_k R_{ik} h_k
    S^T_{ki} = \sum_k R_{ik} S_{ik} h_k \,.
\label{eq:gi1}
\ee
The vectors $(h_1,h_2,h_3)$ and $(g_1,g_2,g_3)$ are linearly
related by
\be
    \left[ \begin{array}{c} g_1 \\ g_2 \\ g_3 \end{array} \right]
    = R \ds S
    \left[ \begin{array}{c} h_1 \\ h_2 \\ h_3 \end{array} \right]
\,,
\ee
where $.*$ denotes the entry-wise multiplication of two matrices,
also known as the Schur product or the Hadamard product.
It is useful to expand $g_i$ in \eq{gi1} explicitly
\be
    g_i = R_{i1} \, S_{i1} \; h_1 + R_{i2} \, S_{i2} \; h_2
        + R_{i3} \, S_{i3} \; h_3
\,.
\ee
Then, we can prove the first group of inequalities
\be
    |g_i| \leq |R_{i1} S_{i1}| \; h_1 + |R_{i2} S_{i2}| \; h_2
         + |R_{i3} S_{i3}| \; h_3
        \leq \max_i h_i = h_1
\label{eq:group1}
\ee
We use the fact that $R,S \in$ SO(3) to prove the second
inequality:
$R,S$ consists of orthonormal rows and columns.
Hence, $(|R_{i1}|,|R_{i2}|,|R_{i3}|)$ and
$(|S_{i1}|,|S_{i2}|,|S_{i3}|)$ are unit vectors, and their inner
product $|R_{i1} S_{i1}| + |R_{i2} S_{i2}| + |R_{i3} S_{i3}| \leq
1$.  We refer to this argument, which is frequently used, as the
``inner product argument''.
The second group of inequalities can be proved by
\be
    \sum_i |g_i| = \sum_i \laL \sum_k R_{ik} S_{ik} \; h_k \raL
    \leq  \sum_k \lpL \sum_i |R_{ik}| |S_{ik}| \rpL |h_k|
    \leq \sum_k h_k = 1
\,.
\label{eq:group2}
\ee
The second inequality in \eq{group2} is due to $\sum_i |R_{ik}|
|S_{ik}| \leq 1$, obtained again by the inner product argument.
This proves all of
\bea
    g_1 + g_2 + g_3 \leq 1
\,,~~
    g_1 - g_2 - g_3 \leq 1
\,,~~
    - g_1 + g_2 - g_3 \leq 1
\,,~~
    - g_1 - g_2 + g_3 \leq 1
\eea
%
%
Finally,
\be
    g_1 + g_2 + g_3 =
    \left( \! \! \begin{array}{r}
    R_{11} \, S_{11} \\ + R_{21} \, S_{21} \\ + R_{31} \, S_{31}
    \end{array} \! \right) h_1
    +
    \left(\! \! \begin{array}{r}
    R_{12} \, S_{12} \\ + R_{22} \, S_{22} \\ + R_{32} \, S_{32}
    \end{array} \! \right) h_2
    +
    \left(\! \! \begin{array}{r}
    R_{13} \, S_{13} \\ + R_{23} \, S_{23} \\ + R_{33} \, S_{33}
    \end{array} \! \right) h_3
    =
    \lambda_1 h_1 + \lambda_2 h_2 + \lambda_3 h_3
\ee
where $\lambda_i$ is the coefficient of $h_i$ in the parenthesis.
The inner product argument implies $|\lambda_i| \leq 1$. Moreover,
we will prove $\sum_i \lambda_i \geq -1$ shortly, which implies
\bea
    g_1 + g_2 + g_3 & \geq &
    \lambda_1 h_1 + \lambda_2 h_2 + (-1 -\lambda_1 -\lambda_2) h_3
\non
\\
    & = & \lambda_1 (h_1 - h_3) + \lambda_2 (h_2 - h_3) - h_3
\non
\\  & \geq & - h_1 - h_2  + h_3 = - (1 - 2 h_3)
\label{eq:group31}
\eea
where \eq{group31} is the minimum of the previous line, attained
at $\lambda_1 = \lambda_2 = -1$ and $\lambda_3 = 1$.
We now prove $\sum_i \lambda_i \geq -1$.
First,
\bea
    \sum_i \lambda_i
    = R_{11} S_{11}  + R_{21} S_{21}  + R_{31} S_{31}
    + R_{12} S_{12}  + R_{22} S_{22}  + R_{32} S_{32}
    + R_{13} S_{13}  + R_{23} S_{23}  + R_{33} S_{33}
    = {\rm tr}(R^T S)
\eea
%
%
%
As $R,S \in$ SO(3), $R^T S \in$ SO(3).
Each SO(3) matrix is a spatial rotation, therefore having the
eigenvalue $+1$ that corresponds to the vector defining the
rotation axis.
Moreover, any SO(3) matrix has determinant $1$.
Therefore, the eigenvalues are generally given by $1$, $e^{i
\phi}$, $e^{-i \phi}$ and the trace is $1 + 2 \cos(\phi) \geq -1$.
This completes the proof of \eq{group31}.
The last $3$ of the $4$ inequalities
\bea
    + g_1 + g_2 + g_3 \geq - (1 - 2 h_3)
\,,~~   + g_1 - g_2 - g_3 \geq - (1 - 2 h_3) \,, \non
\\
    - g_1 + g_2 - g_3 \geq - (1 - 2 h_3)
\,,~~   - g_1 - g_2 + g_3  \geq - (1 - 2 h_3) \,,
\label{eq:group3}
\eea
can be proved similarly.
For example, consider
\be
    g_1 - g_2 - g_3 =
    \left( \!\! \begin{array}{r}
    R_{11} \, S_{11} \\ - R_{21} \, S_{21} \\ - R_{31} \, S_{31}
    \end{array} \! \right) h_1
    +
    \left( \!\! \begin{array}{r}
    R_{12} \, S_{12} \\ - R_{22} \, S_{22} \\ - R_{32} \, S_{32}
    \end{array} \! \right) h_2
    +
    \left( \!\! \begin{array}{r}
    R_{13} \, S_{13} \\ - R_{23} \, S_{23} \\ - R_{33} \, S_{33}
    \end{array} \! \right) h_3
\,.
\ee
The previous argument for $g_1 + g_2 + g_3$ applies by redefining
$R$ to be $\matthree{1}{0}{0}{0}{-1}{0}{0}{0}{-1} \times R$.
\vspace*{1ex}

Altogether, the inequalities in Eqs.~(\ref{eq:group1}),
(\ref{eq:group2}), and (\ref{eq:group3}) satisfied by
$(g_1,g_2,g_3)$ are precisely the defining inequalities for $\P_H$
in \eq{ineq}. Therefore, the diagonal part of any $RD_HS^T$ is in
$\P_H$, and $\C_H = \P_H$.

\subsection{Optimization over $\P_H$}
\label{sec:opsol}

Having proved $\C_H = \P_H$, we can solve the optimal simulation
problem given $D_H$ and $D_{H'}$ by finding the unique
intersection of the semi-line $\lambda D_{H'}$ with the boundary
of $\P_H$ (see Section \ref{sec:glunorm}).  We now explicitly work
out $s_{H|H'}$, i.e. the value of $\lambda$ in the intersection,
as a function of $H$ and $H'$.

Let all the symbols be as previously defined.  The intersection is
given by $\vec{v} = s_{H'|H} (h_1',h_2',h_3')$, so that
\be
    s_{H'|H} = {||\vec{v}||_1 \over ||(h_1',h_2',h_3')||_1}
         = {||\vec{v}||_1 \over h_1'+h_2'+|h_3'|} \,,
\ee
where $||\vec{v}||_1$ for a vector $\vec v$ is the sum of the
absolute values of the entries.
The set $\P_H$ has only $3$ types of boundary faces. Therefore,
there are only $3$ possibilities where the intersection can occur:
\begin{enumerate}
\item
On the group of faces given by $x+y+z = 1, -x+y-z = 1, -x-y+z = 1,
x-y-z = 1$.  In this case, $|| \vec{v} ||_1 = 1$, and $s_{H'|H} =
{1 \over h_1'+h_2'+|h_3'|}$.
\item
On the group of faces $x+y-z = 1-2 h_3, x-y+z = 1-2 h_3, -x+y+z =
1-2 h_3, -x-y-z = 1-2 h_3$. In this case, $|| \vec{v} ||_1 = 1 - 2
h_3$, and $s_{H'|H} = {1 - 2 h_3 \over h_1'+h_2'+|h_3'|}$.
\item
On the group of faces $x=\pm h_1, y = \pm h_1, z = \pm h_1$.
In this case, $\vec{v} = {h_1 \over h_1'} (h_1', h_2', h_3')$
(note $h_1'/h_1 \geq 0$), $|| \vec{v} ||_1 = {h_1 \over h_1'}
(h_1'+h_2'+|h_3'|)$ (not constant on the face), and $s_{H'|H} =
{h_1 \over h_1'}$.
\end{enumerate}
Note that when $H'$ is in normal form, $\vec{v}$ can only fall on
$x+y+z = 1$, $x+y-z = 1-2 h_3$, and $x=h_1$ in each of case 1, 2,
and 3.
We now characterize the $(h_1',h_2',h_3')$ belonging to each case.
\begin{itemize}
\item Case 1.  Note that the face of $\P_H$ on $x+y+z=1$ is the convex
hull of $(h_1,h_2,h_3)$ and all permutations of the entries.  The
hexagon contains exactly all vectors $\vec{v}$ {\em majorized} by
$(h_1,h_2,h_3)$, $\vec{v} \prec (h_1,h_2,h_3)$ (see next section
for definition of majorization).  Hence, $(h_1',h_2',h_3')$ is in
case 1 if and only if it is proportional to some $\vec{v} \prec
(h_1,h_2,h_3)$.\footnote{ The fact $(h_1',h_2',h_3') \prec
(h_1,h_2,h_3)$ is a necessary condition for efficient simulation
is independently proved in Ref.~\cite{Wocjan01}.  }

\item Case 3.  In this case, $\vec{v} = (h_1, {h_1 h_2' \over h_1'},
{h_1 h_3' \over h_1'})$.  Thus $(h_1',h_2',h_3')$ is in case 3 iff
$({h_1 h_2' \over h_1'}, {h_1 h_3' \over h_1'})$ is within the
rectangle with vertices
$(h_2,h_3),(h_3,h_2),(-h_2,-h_3),(-h_3,-h_2)$.
\be \mbox{\psfig{file=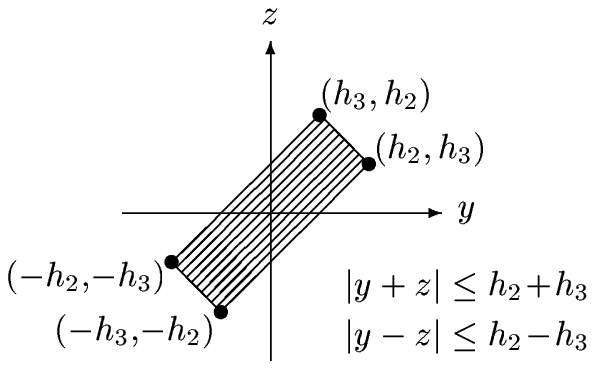,width=1.8in}}
\ee
Hence, $(h_1',h_2',h_3')$ is of case 3
\bea
    {\rm iff} \hspace*{2ex}
    \laL {h_1 h_2' \over h_1'} + {h_1 h_3' \over h_1'} \raL \leq h_2+h_3
& ~~\mbox{and}~~ &
    \laL {h_1 h_2' \over h_1'} - {h_1 h_3' \over h_1'} \raL \leq h_2-h_3
\\
    {\rm iff} \hspace*{10ex}
    {h_1 \over h_2+h_3} \leq {h_1' \over h_2'+h_3'}
& ~~\mbox{and}~~ &
    {h_1 \over h_2-h_3} \leq {h_1' \over h_2'-h_3'}
\label{eq:case3}
\eea

\item Case 2. This contains all $(h_1',h_2',h_3')$ not in case 1 or 3.
\end{itemize}

\noin The intersection on a boundary face can be easily decomposed
as a convex combination of at most $3$ vertices in $P_{24}$.
The decomposition directly translates to an optimal protocol
(using the discussion at the end of Section \ref{sec:glunorm})
with at most 3 types of conjugation.
\vspace*{1ex}

\subsection{Optimal simulation, polyhedron $\P_H$, and s-majorization}
\label{sec:smaj}

The problem of Hamiltonian simulation can also be analyzed from
the perspective of a majorization-like relation, which provides a
compact language to present the main results of this paper.

Let us recall the standard notions of majorization and
submajorization as defined in the space of $n$-dimensional real
vectors.  Let $u$ be an $n$-dimensional vector with real
components $u_i$, $i=1,\cdots,n$. We denote by $u^{\downarrow}$
the vector with components $u_1^{\downarrow} \geq u_2^{\downarrow}
\geq \cdots \geq u_n^{\downarrow}$, corresponding to $|u_i|$
decreasingly ordered.
%
%
Then, for two vectors $u$ and $v$, $u$ is submajorized or weakly
majorized by $v$, written $u \prec_w v$, if
\bea
u_1^{\downarrow} &\leq& v_1^{\downarrow} \,,\\
u_1^{\downarrow}+ u_2^{\downarrow}
& \leq & v_1^{\downarrow}+ v_2^{\downarrow} \,,\\
&\vdots&\\
u_1^{\downarrow} + u_2^{\downarrow} + \cdots u_n^{\downarrow} &
\leq & v_1^{\downarrow}+ v_2^{\downarrow} + \cdots
v_n^{\downarrow} \,.
\eea
In case of equality in the last equation, we say that $u$ is
majorized by $v$, and write $u\prec v$.

These notions can be extended to real matrices.  Suppose $M$ and
$N$ are two $n\times n$ real matrices.  Let sing$(M)$ denote the
set of singular values of the matrix $M$ (and similarly for $N$).
Then, $M \prec_w N$, when sing$(M) \prec_w $ sing$(N)$.
Thus, majorization endows the set of real matrices with a partial
order, and a notion of equivalence,
 \be
 M \sim O_1 M O_2\,,
 \ee
where $O_i \in O(n)$ are orthogonal, and the transformation $M
\rightarrow O_1 M O_2$ preserves the singular values.
A ``convex sum'' characterization of weak majorization
 \be
 M \prec_w N \Leftrightarrow M = \sum_i p_i O_{i1} N O_{i2}
 \ee
also holds~\cite{proofmaj}, meaning that $N$ always weakly
majorizes a (left and right) orthogonal mixing of itself.

We now introduce the notion of special majorization,
s-majorization for short, on $n\times n$ real matrices $M$.
First, we define a new equivalence relation in terms of any $R,S
\in$ SO($n$):
 \be
    M \sim_s RMS \,.
 \label{eq:sequiv}
 \ee
Then, we introduce the s-majorization relation by means of the
``convex sum'' characterization,
 \be
 M \prec_s N \,\Leftrightarrow\, M = \sum_i p_i R_i N S_i \,.
 \label{eq:smajdef1}
 \ee
That is, $M$ is s-majorized by $N$ when $M$ is a (left and right)
special orthogonal mixing of $N$.
Equations (\ref{eq:sequiv}) and (\ref{eq:smajdef1}) suggest
s-ordering for a real vector $u$, $u^{\downarrow s}$:
\be
    (|u|_1^{\downarrow}, |u|_2^{\downarrow}, \cdots,
    |u|_{n-1}^{\downarrow}, {\rm sg}(\Pi_i u_i) |u|_n^{\downarrow})
\,.
\ee
In other words, we rearrange the absolute values of $u_i$ in
decreasing order, and put the sign of the last element to be the
product of all the original signs.
Our results in Section \ref{sec:ceqp}-\ref{sec:opsol} imply, for
$n = 3$, the following characterization of s-majorization:
\begin{quote}
Let $u$ and $v$ be $3$-dimensional vectors, and their s-ordered
versions be $u^{\downarrow_s} = (u_1,u_2,u_3)$ and
$v^{\downarrow_s} = (v_1,v_2,v_3)$. Then $u \prec_s v$ if and only
if
 \bea
 u_1        & \leq & v_1        \,, \non \\
 u_1+u_2-u_3    & \leq & v_1+v_2-v_3    \,, \non \\
 u_1+u_2+u_3    & \leq & v_1+v_2+v_3    \,.
 \label{eq:sm}
 \eea
\end{quote}

\noindent{\bf Remark}~ Note that $u \prec_s v \Rightarrow u
\prec_w v$.
Moreover, when sg$(\Pi_i u_i)=$ sg$(\Pi_i v_i)$ and $\sum_i
u_i=\sum_i v_i$, then $\prec_w$, $\prec_s$, and $\prec$ are all
equivalent.
\vspace*{1ex}

We can now state our result in Hamiltonian simulation in the
language of s-majorization:
\begin{quote}
{\bf Theorem:} Let $H = \sum_i h_i \sigma_i \ot \sigma_i$ and $H'
= \sum_i h_i' \sigma_i \ot \sigma_i$, $h = (h_1,h_2,h_3)$, and $h'
= (h_1',h_2',h_3')$.  Then
\be
    H' \leq_{\rm LU} H ~~\Leftrightarrow~~ h' \prec_s h.
\ee
The optimal simulation factor is given by $s_{H'|H} = \max_{s h'
\prec_s h} s$.

{\em Proof}: Note that when both $h$ and $h'$ are s-ordered,
\eq{ineq} that characterizes $\P_H$ reduces to
 \bea
 h_1'           & \leq & h_1            \,, \non \\
 h_1'+h_2'-h_3'     & \leq & h_1+h_2-h_3        \,, \non \\
 h_1'+h_2'+h_3'     & \leq & h_1+h_2+h_3        \,.
 \label{eq:sm2}
 \eea
which is the condition for s-majorization. Thus $h' \in \P_H$ iff
$h' \prec_s h$, and $H' \leq_{\rm LU} H ~\Leftrightarrow~ h'
\prec_s h$. $\Box$
\end{quote}

\section{Hamiltonian Simulation with LU$+$anc}
\label{sec:luanc}

In this section we will show that the use of uncorrelated ancillas
does not help when simulating one two-qubit Hamiltonian with
another, so that all results on efficient and optimal simulation
under LU hold under LU$+$anc.
We prove this by describing the most general LU$+$anc protocol and
reducing it to an LU protocol.
\vspace*{1ex}

In this scenario, qubits $A$ and $B$ are respectively appended
with ancillas $A'$ and $B'$, which have finite but arbitrary
dimensions.
The initial state of $A'B'$ can be chosen to be a pure product
state $\ket{0_{A'}} \ot \ket{0_{B'}}$.
At the final stage of the simulation, the ancillas $A'$ and $B'$
may be correlated, but $A'B'$ is uncorrelated with $AB$ if the
latter is to evolve unitarily according to $H'$.
The local unitary transformations $U_i$ and $V_i$ can act on $AA'$
and $BB'$ respectively.  This feature distinguishes LU$+$anc from
LU.
\vspace*{1ex}

The most general LU$+$anc protocol to simulate $H'$ with $H$ can
be described as
\bea
        \lpL U_f \ot V_f \times U_n \ot V_n \; e^{-i H t_n} \;
    U_n^\dag \ot V_n^\dag
        \times \cdots \times
        U_1 \ot V_1 \; e^{-i H t_1} \;
        U_1^\dag \ot V_1^\dag \rpL \; |\psi\> \ot |0_{A'}\> \ot |0_{B'} \>
    \hspace*{20ex}
\non
\\
    \hspace*{50ex} = \lpm e^{-iH't'} |\psi\> \rpm
    \ot \lpm W_{A'B'}(t_1,\cdots,t_n) |0_{A'}\> \ot |0_{B'}\> \rpm
    \,.
\label{eq:gensim2}
\eea
In Appendix \ref{sec:derivelinform} we have shown that for
infinitesimal times \eq{gensim2} leads to
\be
    sH'_{AB} =  \<0_{A'}| \ot \<0_{B'}| \lbL
    \sum_k p_k \, U_k \ot V_k \, (H \ot I_{A'B'}) \, U_k^\dag \ot V_k^\dag
    \rbL |0_{A'}\> \ot |0_{B'}\>,
\label{eq:finalanc2}
\ee
where $p_k\equiv t_k/t$ and $s\equiv t'/t$.  Let $M_k \equiv
\bra{0_{A'}}U_k$ and $N_k\equiv \bra{0_{B'}}V_k$. We can write
\eq{finalanc2} as
\be
    sH' = \sum_k p_k \, M_k \ot N_k \, (H \ot I_{A'B'})
    \, M_k^\dag \ot N_k^\dag
\,.
\label{eq:ancillas4}
\ee
Note that this is the LU$+$anc analogue of \eq{cosumH} for LU.  In
this case, $H$ is replaced by $H \ot I_{A'B'}$ and the local
unitaries are replaced with more general transformations.
\vspace*{1ex}

We focus on just one term in the convex combination of
\eq{ancillas4}, $M \ot N \, (H_{AB} \ot I_{A'B'}) \, M^\dag \ot
N^\dag$, with $M \equiv \<0_{A'}|U_k$ and $N \equiv \<0_{B'}|V_k$.
We will show how to obtain the same contribution to $H'$ using
only local unitaries on $A$ and $B$ to establish the equivalence
of LU and LU$+$anc.  First, note that
\be
    M \ot N \, (H_{AB} \ot I_{A'B'}) \, M^\dag \ot N^\dag
    = \E_A \circ \E_B (H)
\label{eq:effonH}
\ee
where $\E_A(\tau) \equiv M (\tau \otimes I_{A'}) M^\dag$ and
similarly for $\E_B$.
We emphasize that $\E_{A,B}$ are linear operators on matrices that
are {\em not} necessarily quantum operations~\cite{Nielsen00},
despite various resemblances to the latter.
One can check that $\E_A$ is {\em unital}, i.e.~$\E_A(I)=I$, by
using $M = \bra{0_{A'}} U$.
Furthermore, $\E_A$ is {\em completely positive}~\cite{Nielsen00},
because an operator-sum representation $\E_A (\tau) \equiv \sum_i
F_i \tau F_i^\dagger$ can be obtained by expanding $I_{A'}$ in
terms of some basis $\{\ket{i_{A'}}\}$, and by writing $F_i = M
|i_{A'}\> = \<0_{A'}| U|i_{A'}\>$.
However, in general, $\E_A$ is neither trace nonincreasing or
trace nondecreasing, though $\tr_A \sum_i F_i^\dag F_i = \tr_A
\sum_i \<i_{A'}|U^\dag |0_{A'}\>\<0_{A'}| U|i_{A'}\> = 2$.
For each $F_i$, we can obtain the singular value decomposition
$\F_i = W_{2i} Q_i W_{1i}$, where $W_{1i}$ and $W_{2i}$ are
unitary, and
\be
    Q_i = \left[ \begin{array}{cc} q_{i1} & 0 \\
                           0   & q_{i2} \end{array} \right]
\ee
is diagonal and positive semidefinite.
Altogether,
\be
    \E_A (\tau) = \sum_i {1 \over 2} (q_{i1}^2 + q_{i2}^2) \;
    W_{2i} \; \tilde{Q}_i \; W_{1i} \; \tau \;
    W_{1i}^\dag \; \tilde{Q}_i \; W_{2i}^\dag
\ee
where
$
    \tilde{Q}_i = \sqrt{2} \left[ \begin{array}{cc} \cos \theta_i & 0 \\
                           0   & \sin \theta_i \end{array} \right]
$
and $\cos \theta_i = q_{i1}/\sqrt{q_{i1}^2 + q_{i2}^2}$.
We now claim that we can replace the action of $\tilde{Q}_i$ by
${\cal Q}_i(\tau) = (1 - \cos \theta \sin \theta) \; I \tau I +
\cos \theta \sin \theta \; \sigma_z \tau \sigma_z$, i.e.,
replacing $\E_A$ by the following:
\be
    \tilde{\E}_A (\tau) = \sum_i {1 \over 2} (q_{i1}^2 + q_{i2}^2) \;
    W_{2i} \; {\cal Q}_i(W_{1i} \; \tau \; W_{1i}^\dag) \; W_{2i}^\dag
\ee
It is straightforward to verify that
\bea
    & & \tilde{Q}_i I \tilde{Q}_i = I + \cos 2 \theta \; \sigma_z \,,~~
    \tilde{Q}_i \sigma_x \tilde{Q}_i = \sin 2 \theta \; \sigma_x \,,~~
    \tilde{Q}_i \sigma_y \tilde{Q}_i = \sin 2 \theta \; \sigma_y \,,~~
    \tilde{Q}_i \sigma_z \tilde{Q}_i = \cos 2 \theta \; I + \sigma_z \,.
\label{eq:tq}
\\
    & & {\cal Q}_i(I) = I \,,~~
    {\cal Q}_i(\sigma_x) = \sin 2 \theta \; \sigma_x \,,~~
    {\cal Q}_i(\sigma_y) = \sin 2 \theta \; \sigma_y \,,~~
    {\cal Q}_i(\sigma_z) = \sigma_z \,.
\label{eq:cq}
\eea
As $H$ is purely nonlocal, the input to $\E_A$ in \eq{effonH} is
always traceless.  Now, consider how $\E_A$ and $\tilde{\E}_A$ act
on $\sigma_{x,y,z}$.  The only difference is that ${\cal Q}_i$ is
not producing the extra $I$ component produced by $\tilde{Q}_i
\sigma_z \tilde{Q}_i$.  This extra $I$ component has a final
contribution as local terms in \eq{effonH}, which can be ignored.
Finally, we note that $\sum_i {1 \over 2} (q_{i1}^2 + q_{i2}^2) =
{1 \over 2} \sum_i \tr F_i^\dag F_i = 1$, so that $\tilde{\E}_A$
is indeed a convex combination of the individual terms, each in
turns a mixture of unitary operations on $A$.
Applying the same argument to $\E_B$, Alice and Bob only need to
perform local unitaries in the simulation step of \eq{effonH}.

\section{Discussion}
\label{sec:discussion}

First, we point out that the normal form for Hamiltonians acting on
$2$ qubits (Sec.~\ref{sec:normform22}) is {\em symmetric} with
respect to exchanging the systems $A$ and $B$.  More formally, let
${\cal S}(M_1 \ot M_2) = M_2 \ot M_1$ be the (nonlocal) {\sc swap}
operation.  Then $H \equiv_{\rm LU} {\cal S}(H)$.  This
has important consequence -- any task generated by the Hamiltonian can
be done equally well with the role of Alice and Bob interchanged.
\vspace*{1ex}

In higher dimensions, the property $H \equiv_{\rm LU} {\cal S}(H)$ no
longer holds.  For example, $H \not\leq_{\rm LU} {\cal S}(H)$ and
${\cal S}(H) \not\leq_{\rm LU} H$ for the Hamiltonian (see
Ref.~\cite{noswap} for proof):
\be
	H =  
	\left[ \begin{array}{ccc} 
	1 & 0 & 0  \\ 
	0 & -1 & 0  \\
	0 & 0 & 0 \end{array} \right] \ot
 	\left[ \begin{array}{ccc} 
	1 & 0 & 0  \\ 
	0 & 1 & 0  \\
	0 & 0 & -2 \end{array} \right] 
\label{eq:noswap}
\ee
In fact, if $H = H_1 \ot H_2$ where $H_1$ and $H_2$ are members of a
traceless orthogonal basis with different eigenvalues, ${\cal S}(H)
\not\leq_{LU} H$ and $H \not\leq {\cal S}(H)$.
This also has important consequences -- in higher dimensions, the
nonlocal degrees of freedom of a Hamiltonian cannot be characterized
by quantities that are symmetric with respect to $A$ and $B$, such as
eigenvalues of $H$ (independently reported in Ref.~\cite{Chen01s}).
Any normal form necessarily contains terms of the form $c_{ij} \eta_i
\ot \eta_j$ for some nonzero $c_{ij}$ and the matrix with entries 
$c_{ij}$ {\em cannot} be symmetric. 
\vspace*{1ex}

Second, we revisit the notion of efficiency in Hamiltonian simulation.
Our definition of $H' \leq H$ depends on the normalization of both $H$
and $H'$.  One method to remove the normalization dependence is to
require $h_1'+h_2'+|h_3'| = 1$.  Alternatively, we can consider the
product $s_{H|H'} s_{H'|H}$, that measures the inefficiency of
interconverting $H$ and $H'$ independent of the normalization of the
Hamiltonians.  We found that (proof omitted) when $h_3' \geq 0$,
$s_{H|H'} s_{H'|H} \geq {1 \over 3}$.  Otherwise, 
$s_{H|H'} s_{H'|H} \geq {1 \over 9}$, with the lower bound attained 
at $h = (1/3,1/3,1/3)$ and $h'=(1/3,1/3,-1/3)$.
\vspace*{1ex}

Third, we have considered the optimal simulation of one two-qubit
Hamiltonian using another, both arbitrary but known.  We can apply
the characterization of $\P_H$ to analyze other interesting
problems.
For example, inverting a known Hamiltonian is equivalent to
setting $H' = -H$.  Without loss, assume $h_3 \geq 0$ and
$h_1+h_2+h_3 = 1$. Using the analysis in Section \ref{sec:opsol},
the intersection is of case 2.
%
%
Therefore, $s_{-H|H} = -(1-2h_3)$.
The worst case is inverting ${1 \over 3}(\sigma_x \ot \sigma_x +
\sigma_y \ot \sigma_y + \sigma_z \ot \sigma_z)$ in which case
$s_{-H|H} = 1/3$.
In contrast, any protocol for inverting an unknown Hamiltonian can
invert the worst known Hamiltonian, thus $s_{-H|H} \leq 1/3$. This
is achievable using the following protocol:
\be
    \sigma_x \ot I \;\; e^{-iHt'} \; (\sigma_x
    \sigma_y \ot I) \;\; e^{-iHt'} \; (\sigma_y
    \sigma_z \ot I) \;\; e^{-iHt'} \; \sigma_z \ot I = e^{-i(-H)t'/3} \,.
\ee
We can also improve on the time requirement for simulation
protocols for $n$-qubit pairwise coupling
Hamiltonians~\cite{Dodd01} with our construction.  Instead of
selecting a term by term simulation using a single nonlocal Pauli
operator acting on a pair of qubit, one can directly simulate the
desired coupling between the pair with any given one in a time
optimal manner.

\section{Conclusion}
\label{sec:conclusion}

We have discussed various notions of Hamiltonian simulation.
Focusing on dynamics simulation, we show its equivalence to
infinitesimal simulation, and the intrinsic time independence of
the protocols. We also show the possibility of simulating one
nonlocal Hamiltonian with another without ancillas in any two
$d$-dimensional system. Our main results are on two-qubit
Hamiltonians, in which case, for any Hamiltonian $H$, we
characterize all $H'$ that can be simulated efficiently, and
obtain the optimal simulation factor and protocol.  We obtain our
results by considering a simple polyhedron that is related to some
majorization-like relations.  Our results show that the two-qubit
Hamiltonians are endowed with a partial order, in close analogy to
the partial ordering of bipartite pure states under local
operations and classical communication~\cite{Nielsen98}.
\vspace*{1ex} 

We have restricted our attention to simulation protocols that are
infinitesimal, one-shot, deterministic, and without the use of
entangled ancillas and classical communication.  We also restricted
our attention to bipartite systems.
Extensions to the unexplored regime, and alternative direction 
such as other nonlocal tasks will prove useful, and are being 
actively pursued.  

\section{Acknowledgements}
\label{sec:ack}

We thank Wolfgang D\"ur, John Smolin, and Barbara Terhal for
suggestions and discussions that were central to this work, as
well as Herbert Bernstein, Ben Recht, and Aram Harrow for
additional helpful discussions and comments.
CHB and DWL are supported in part by the NSA and ARDA under the US
Army Research Office, grant DAAG55-98-C-0041.
GV is supported by the European Community project EQUIP (contract
IST-1999-11053), and contract HPMF-CT-1999-00200.


\appendix

\section{Notions of simulation}
\label{sec:simnots}

We consider various notions of using a Hamiltonian $H$ to simulate
the evolution due to $H'$ for time $t'$.

In dynamics simulation, the evolution of the system is close to
$e^{-i H' t''}$ after an operation time of $\mu t''$ for constant
$\mu$ and $\forall~ t'' \in [0,t']$.  It is possible to relax this
requirement, so that, $\mu(t'')$ is a function of $t''$, and
without loss of generality, $\mu(t'')$ is nondecreasing.  We call
this ``variable rate dynamics simulation''.  Finally, in gate
simulation, the only requirement is that, the final evolution is
given by $e^{-i H' t'}$.

As an analogy, let $H'$ be driving along a particular highway from
my house to your house at 100 km/hr. Dynamics simulation is like
driving, biking or walking along the same highway at any {\em
constant} speed. Variable rate dynamics simulation is like driving
along the highway at variable speed, for example, when there is
stop-and-go traffic. The vehicle is always on the trajectory
defined by $H'$. Finally, gate simulation is like going from my
house to your house by any means, for example using local roads,
or flying a helicopter.

It is important to note the difference between dynamics simulation
(or infinitesimal simulation) and variable rate dynamics
simulation.  For example, iterating infinitesimal simulations to
perform dynamics simulation, the ancillas are implicitly discarded
after each iteration, and new ones be used next.  However, it is
possible in variable dynamics simulation that used ancillas can
subsequently be used to accelerate the simulation.  Such phenomena
are known in entanglement generation~\cite{Dur00}.  The more
complicated analysis for variable dynamics simulation will be
addressed in future work.

\section{Simulating zero Hamiltonian in $d \times d$ without ancillas}
\label{sec:viola}

In Ref.~\cite{Viola99}, it is shown that for any $d$-dimensional
square matrix $M$,
\be
    \sum_{ij} U_{ij} M U_{ij}^\dag
    = \tr{M} \; {I \over d}
\ee
where
\be
    U_{ij} = \left[
    \begin{array}{ccccc}
    1 & 0 & 0 & 0 & 0 \\
    0 & \omega & 0 & 0 & 0 \\
    0 & 0 & \omega^2 & 0 & 0 \\
    0 & 0 & 0 & \cdot & 0 \\
    0 & 0 & 0 & 0 & \omega^{d-1} \end{array} \right]^i
    \left[ \begin{array}{ccccc}
    0 & 1 & 0 & 0 & 0 \\
    0 & 0 & 1 & 0 & 0 \\
    0 & 0 & 0 & 1 & 0 \\
    0 & 0 & 0 & 0 & 1 \\
    1 & 0 & 0 & 0 & 0 \end{array} \right]^j
\ee
and $\omega$ is a primitive $d$-th root of unity.
$H$ can simulate ${\bf 0}$ using the protocol:
\be
    \Pi_{ij} (U_{ij} \ot I) \; e^{-i H t} \; (U_{ij}^\dag \ot I) \approx
    e^{-i \sum_{ij} (U_{ij} \ot I) H (U_{ij}^\dag \ot I) t} =
    e^{-i I \ot K_B  t}
\ee
which is local and can be removed.

\section{Arbitrary Hamiltonian simulation in $d \times d$ without ancillas}
\label{sec:ddimsim}

Let $H$ and $H'$ act on two $d$-dimensional systems.
We use the following (nonorthonormal) basis for traceless
hermitian operators acting on a $d$-dimensional system:
\bea
    & &
    \eta_1 = \left[ \begin{array}{rrrr}
        1 & 0 & 0 & ... \\
        0 & -1 & 0 & ... \\
        0 & 0 & 0 & ... \\
        \vdots & \vdots & \vdots & \vdots
        \end{array} \right] \,,~~
    \eta_2 = \left[ \begin{array}{rrrr}
        1 & 0 & 0 & ... \\
        0 & 0 & 0 & ... \\
        0 & 0 & -1 & ... \\
        \vdots & \vdots & \vdots & \vdots
        \end{array} \right] \,,~~
    \cdots ~~\,,~~
    \eta_{d-1} = \left[ \begin{array}{rrrr}
        1 & 0 & 0 & ... \\
        0 & 0 & 0 & ... \\
        0 & 0 & 0 & ... \\
        \vdots & \vdots & \vdots & -1
        \end{array} \right] \,,~~
\non
\\
    & &
    \eta_d = \left[ \begin{array}{rrrr}
        0 & 1 & 0 & ... \\
        1 & 0 & 0 & ... \\
        0 & 0 & 0 & ... \\
        \vdots & \vdots & \vdots & \vdots
        \end{array} \right] \,,~~
    \eta_{d+1} = \left[ \begin{array}{rrrr}
        0 & -i & 0 & ... \\
        i & 0 & 0 & ... \\
        0 & 0 & 0 & ... \\
        \vdots & \vdots & \vdots & \vdots
        \end{array} \right] \,,~~
    \eta_{d+2} = \left[ \begin{array}{rrrr}
        0 & 0 & 1 & ... \\
        0 & 0 & 0 & ... \\
        -1 & 0 & 0 & ... \\
        \vdots & \vdots & \vdots & \vdots
        \end{array} \right] \,,~~
    \eta_{d+3} = \left[ \begin{array}{rrrr}
        0 & 0 & -i & ... \\
        0 & 0 & 0 & ... \\
        i & 0 & 0 & ... \\
        \vdots & \vdots & \vdots & \vdots
        \end{array} \right] \,,~~
    \cdots ~~\,,~~
\non
\eea
Let $H = \sum_{ij} c_{ij} \, \eta_i \ot \eta_j$ and $H' =
\sum_{ij} c_{ij}' \eta_i \ot \eta_j$.  To show that $sH' \leq_{\rm
LU} H$ for some $s > 0$, it suffices to show that $s \; \eta_1 \ot
\eta_1 \leq_{\rm LU} H$, since $\eta_1 \ot \eta_1 \equiv_{\rm LU}
\pm \eta_i \ot \eta_j$ for all $i,j$.
Furthermore, $\eta_1 \ot \eta_1$ can be simulated if one can
simulate $|i\>\<i| \ot |j\>\<j|$ and $-|i'\>\<i'| \ot |j'\>\<j'|$
for any $i,j,i',j'$.
\vspace*{1ex}

Without loss of generality, $c_{11} \neq 0$.
We first use $H$ to simulate its diagonal components, $H_d =
\sum_{i,j = 1}^{d-1} c_{ij} \, \eta_i \ot \eta_j$:
\be
    H_d = {1 \over d^2} \sum_{i,j=0}^{d-1}
    \left[ \begin{array}{rrrr}
        1 & 0 & 0 & ... \\
        0 & \omega & 0 & ... \\
        0 & 0 & \omega^2 & ... \\
        \vdots & \vdots & \vdots & \vdots
    \end{array} \right]^i \! \!
    \ot
    \left[ \begin{array}{rrrr}
        1 & 0 & 0 & ... \\
        0 & \omega & 0 & ... \\
        0 & 0 & \omega^2 & ... \\
        \vdots & \vdots & \vdots & \vdots
    \end{array} \right]^j
    ~ H ~\;
    \left[ \begin{array}{rrrr}
        1 & 0 & 0 & ... \\
        0 & \omega & 0 & ... \\
        0 & 0 & \omega^2 & ... \\
        \vdots & \vdots & \vdots & \vdots
    \end{array} \right]^{i\dag} \!  \!
    \ot
    \left[ \begin{array}{rrrr}
        1 & 0 & 0 & ... \\
        0 & \omega & 0 & ... \\
        0 & 0 & \omega^2 & ... \\
        \vdots & \vdots & \vdots & \vdots
    \end{array} \right]^{j\dag}
\ee
$H_d$ can further be used to simulate $c_{11} |2\>\<2| \ot
|2\>\<2|$, using the protocol
\be
    {1 \over (d-1)^2} \sum_{i,j=0}^{d-2}
    \left[ \begin{array}{rrrrr}
        0 & 0 & 1 & 0 & ... \\
        0 & 1 & 0 & 0 & ... \\
        0 & 0 & 0 & 1 & ... \\
        \vdots & \vdots & \vdots & \vdots & 1 \\
        1 & 0 & 0 & 0 & 0 \\
    \end{array} \right]^i \! \!
    \ot
    \left[ \begin{array}{rrrrr}
        0 & 0 & 1 & 0 & ... \\
        0 & 1 & 0 & 0 & ... \\
        0 & 0 & 0 & 1 & ... \\
        \vdots & \vdots & \vdots & \vdots & 1 \\
        1 & 0 & 0 & 0 & 0 \\
    \end{array} \right]^j
    ~ H_d ~\;
    \left[ \begin{array}{rrrrr}
        0 & 0 & 1 & 0 & ... \\
        0 & 1 & 0 & 0 & ... \\
        0 & 0 & 0 & 1 & ... \\
        \vdots & \vdots & \vdots & \vdots & 1 \\
        1 & 0 & 0 & 0 & 0 \\
    \end{array} \right]^{i\dag} \!  \!
    \ot
    \left[ \begin{array}{rrrrr}
        0 & 0 & 1 & 0 & ... \\
        0 & 1 & 0 & 0 & ... \\
        0 & 0 & 0 & 1 & ... \\
        \vdots & \vdots & \vdots & \vdots & 1 \\
        1 & 0 & 0 & 0 & 0 \\
    \end{array} \right]^{j\dag}
\ee
This corresponds to Alice and Bob each applies an averaging over
all the computation basis states except for $|2\>$.
Since all $\eta_{i \neq 1}$ are traceless on the subspace spanned
by $|i \neq 2\>$, they vanish after the averaging, leaving only a
contribution by $\eta_{1} \ot \eta_1$: \be
    c_{11} \left[ \begin{array}{rrrrr}
        {1 \over d-1} & 0 & 0 & 0 & ... \\
        0 & -1 & 0 & 0 & ... \\
        0 & 0 & {1 \over d-1} & 0 & ... \\
        \vdots & \vdots & \vdots & \vdots & 0 \\
        0 & 0 & 0 & 0 & {1 \over d-1} \\
    \end{array} \right]^{i\dag} \!  \!
    \ot
    \left[ \begin{array}{rrrrr}
        {1 \over d-1} & 0 & 0 & 0 & ... \\
        0 & -1 & 0 & 0 & ... \\
        0 & 0 & {1 \over d-1} & 0 & ... \\
        \vdots & \vdots & \vdots & \vdots & 0 \\
        0 & 0 & 0 & 0 & {1 \over d-1} \\
    \end{array} \right]^{i\dag} \!  \!
\ee
which is equivalent to $c_{11} |2\>\<2| \ot |2\>\<2|$ up to local
terms.
It remains to obtain a term with sign opposite to $c_{11}$.
If some $c_{kl} \neq 0$ has a sign opposite to $c_{11}$, we can
simply repeat the same procedure, with Alice applying an averaging
over all $|i \neq k\>$ and Bob applying an averaging over all $|j
\neq l\>$.
If all $c_{ij}$ has the same sign, Alice can apply an averaging
over all $|i \neq 1\>$ and Bob can apply an averaging over all $|j
\neq 2\>$ to obtain $ - \sum_{i=1}^{d-1} c_{i1} |1\>\<1| \ot
|2\>\<2|$, completing the proof.

\section{Infinitesimal simulation and time independence}
\label{sec:derivelinform}

The most general simulation protocol of $H'$ with $H$ using
LU$+$anc can be described by
\bea
    \hspace*{-10ex}
        \lpm U_f\otimes V_f \times U_n \ot V_n \; e^{-i H t_n}
    \; U_n^\dag \ot V_n^\dag
        \cdots
        U_1 \ot V_1 \; e^{-i H t_1} \; U_1^\dag\ot V_1^\dag\rpm
    \lpm |\psi\> \ot |0_{A'}\> \ot |0_{B'}\> \rpm
\non
\\
    = \lpm e^{-iH't'} |\psi\> \rpm\ot   \lpm
        W_{A'B'} \;|0_{A'}\> \ot |0_{B'}\> \rpm,
    \hspace*{-10ex}
\label{eq:gensim10}
\eea
where the equality must hold for all possible states $\ket{\psi}$
of system $AB$. Here the unitaries $U_i$ and $V_i$, acting on
$AA'$ and $BB'$ respectively, and the partition $\{t_i\}$ of the
time interval $t=\sum_it_i$, correspond to all the degrees of
freedom available for the simulation of $H'$ for time $t'$.  The
initial states of the ancillas $A'$ and $B'$ is $|0_{A'}\> \ot
|0_{B'}\>$, and $W_{A'B'}$ is their residual, unitary evolution,
which is determined by the other degrees of freedom and may create
entanglement between $A'$ and $B'$.

We have argued earlier that optimal dynamics simulation can always
be achieved by a protocol for simulating infinitesimal evolution
times $t'$.
This also implies $t$ being infinitesimal.
Recall that $p_i \equiv t_i/t$ and $s \equiv t'/t$.  We can expand
\eq{gensim10} to first order in $t$ to obtain
\bea
    U_f \ot V_f \times
    \lbL I - i t \sum_i p_i \,
    U_i \ot V_i \; ( H \otimes I_{A'B'} ) \; U_i^\dag \ot V_i^\dag \rbL
    \ket{0_{A'}}\otimes \ket{0_{B'}}
    = \lpm I-itsH' \rpm \otimes \lpm W_{A'B'} \; \ket{0_{A'}}\otimes
    \ket{0_{B'}}\rpm \,.
\label{eq:gensim11}
\eea
The validity of \eq{gensim10} for all $\ket{\psi}$ is used to
obtain \eq{gensim11}, each term of which is taken to be an
operator on $AB$.
It follows from \eq{gensim11} that
\be
    \lpm U_f\ket{0_{A'}}\rpm \otimes \lpm V_f \ket{0_{B'}} \rpm
    = I_{AB} \otimes \lpm W_{A'B'} \ket{0_{A'}}\otimes \ket{0_{B'}} \rpm
    + {\cal O}(t) \,,
\ee
which implies that
\bea
\label{eq:anca}
U_f\, \ket{0_{A'}} &=& I_A\otimes \lpm W_{A'} \,\ket{0_{A'}}\rpm + \O(t),\\
V_f\, \ket{0_{B'}} &=& I_B\otimes \lpm W_{B'} \,\ket{0_{B'}}\rpm + \O(t),\\
W_{A'B'}\;\ket{0_{A'}}\otimes \ket{0_{B'}} &=& \lpm W_{A'} \,
\ket{0_{A'}} \rpm \otimes \lpm W_{B'} \, \ket{0_{B'}}\rpm + \O(t)
\,.
\label{eq:ancfirst}
\eea
Equation (\ref{eq:ancfirst}) implies $W_{A'B'}$ is a product
operator to zeroth order in $t$.
Redefining $U_f$ and $V_f$ is necessary, we can assume
$W_{A'}=I_{A'}$ and $W_{B'}=I_{B'}$.  Explicitly writing down the
most general ${\cal O}(t)$ terms in
Eqs.~(\ref{eq:anca})-(\ref{eq:ancfirst}), we obtain
\bea
    U_f\, \ket{0_{A'}} &=&
    \lpm I_{AA'} - it K_{AA'}\rpm \ket{0_{A'}} + \O(t^2) \,,
\label{eq:uffirst}
\\
    V_f\, \ket{0_{B'}} &=&
    \lpm I_{BB'} - it K_{BB'}\rpm \ket{0_{B'}} + \O(t^2),\,,
\\
    W_{A'B'} \; \ket{0_{A'}}\otimes \ket{0_{B'}}
    &=& \lpm (I_{A'}-itK_{A'}) \ket{0_{A'}} \rpm \ot
            \lpm (I_{B'}- itK_{B'}) \ket{0_{B'}} \rpm
        -it K_{A'B'} \,|0_{A'}\> \ot |0_{B'}\> + \O(t^2).
\label{eq:wabfirst}
\eea
where the unitarity of the operators on the LHS implies the
hermiticity of $K_{AA'}$, $K_{BB'}$, $K_{A'}$, $K_{B'}$, and
$K_{A'B'}$. Substituting
Eqs.~(\ref{eq:uffirst})-(\ref{eq:wabfirst}) in \eq{gensim11}
implies
\bea
    &&\hspace*{-3ex} sH' \otimes
    \lpm I_{A'B'}\ket{0_{A'}}\otimes \ket{0_{B'}}\rpm =
\nonumber
\\
    &&\hspace*{3ex} \lpm \sum_i p_i
    U_i\otimes V_i (H \otimes I_{A'B'}) U_i^{\dagger}\otimes V_i^{\dagger}
    + K_{AA'} + K_{BB'} - K_{A'} - K_{B'} - K_{A'B'} \rpm
    \ket{0_{A'}} \otimes \ket{0_{B'}} + \O(t) \,.
\eea
Projecting this equation on the left onto
$\bra{0_{A'}}\otimes\bra{0_{B'}}$, the terms $K_{AA'} + K_{BB'} -
K_{A'} - K_{B'} - K_{A'B'}$ become local or identity terms.
Taking into account that $H'$ has zero trace and no local terms
(recall Section I.C.1), their contributions vanish, and we obtain
\be
    sH' =  \<0_{A'}| \ot \<0_{B'}| \lbL \sum_i \;p_i
    \; U_i \ot V_i \; (H \ot I_{A'B'}) \; U_i^{\dagger} \ot V_i^{\dagger}
    \rbL |0_{A'}\> \ot |0_{B'}\>
\,.
\label{eq:finalanc}
\ee
In the case we do not have ancillary systems, $U_i$ and $V_i$ only
acts on $A$ and $B$, and \eq{finalanc} reads
\be
    sH' = \sum_i \;p_i\;
    U_i \ot V_i \; H \; U_i^{\dagger} \ot V_i^{\dagger}
\,,
\label{eq:finalnoanc}
\ee
In Eqs.~(\ref{eq:finalanc}) and (\ref{eq:finalnoanc}), the
dependence of the equation on the original infinitesimal times $t$
and $t'$ is only through $s=t'/t$.
This implies any protocol for $t$ and $t'$, applies to $at$ and
$at'$ within the infinitesimal regime.  Thus the protocol, namely,
the set $\{U_i,V_i,p_i\}$ can be considered being independent of
$t$ in the infinitesimal regime.



\begin{thebibliography}{10}
\bibitem{noswap} Let $H = H_1 \ot H_2$, $H_1, H_2$ be members of
a traceless orthogonal basis (under the trace norm) with different
eigenvalues.  We show that ${\cal S}(H) \not\leq_{\rm LU} H$, which 
also implies $H \not\leq_{\rm LU} {\cal S}(H)$.  
%
We can renormalize both $H$ and ${\cal S}(H)$ so that the basis is
orthonormal.
%
The condition ${\cal S}(H) \leq_{\rm LU} H$ implies  
%
\bea
	H_2 \ot H_1 = \sum_i p_i (U_i H_1 U_i^\dag) \ot (V_i H_2 V_i^\dag)
\non
\eea
%
for some unitary $U_i, V_i$.  
%
Multiplying the product of the above with $H_2 \ot H_1$ and taking the 
trace,
%
\bea
\tr (H_2^2) \tr(H_1^2) 
= \sum_i p_i \tr (H_2 U_i H_1 U_i^\dag) \tr (H_1 V_i H_2 V_i^\dag)
\non
\eea
%
However, 
$\tr(H_2 U_i H_1 U_i^\dag) \leq \tr(H_2^2)$ and 
$\tr(H_1 V_i H_2 V_i^\dag) \leq \tr(H_1^2)$, with equalities hold only 
when $U_i H_1 U_i^\dag = H_2$ and $V_i H_2 V_i^\dag = H_1$, implying  
%
$H_1$ and $H_2$ need to have the same eigenvalues. 
%
It is well known that, in $d$-dimensions, the traceless matrices  
%
${\rm diag}(1,1,\cdots,1,-k,0,0,\cdots,0)$ are part of a traceless
orthogonal basis for hermitian $d \times d$ matrices (for example,
the Gell-Mann matrices) completing the proof.
\end{thebibliography}
\end{document}